\documentclass[aps,prc,preprint,amsmath,amssymb,showpacs,preprintnumbers,superscriptaddress]{revtex4-1}
\usepackage{CJK}
\usepackage{graphicx}% Include figure files
\usepackage{dcolumn}% Align table columns on decimal point
\usepackage{bm}% bold math
\usepackage{slashed}
\usepackage{ulem}
\usepackage{epstopdf}
\usepackage{bbold}
\usepackage{mathrsfs}   % added by S.B. Wang to use \mathscr{L}_{N\pi}
\usepackage{color,xcolor}  % added by S.B. Wang to use \color{red}
\usepackage{multirow}  % added by S.B. Wang to use multirow in table
\usepackage{booktabs} % added by S.B. Wang to use \cmidrule(r){5-6}

\usepackage{hyperref}% add hypertext capabilities
\newcommand{\bbm}{\begin{bmatrix}}
\newcommand{\ebm}{\end{bmatrix}}
\newcommand{\bBm}{\begin{Bmatrix}}
\newcommand{\eBm}{\end{Bmatrix}}
\newcommand{\bpm}{\begin{pmatrix}}
\newcommand{\epm}{\end{pmatrix}}

\begin{document}
%\begin{CJK*}{UTF8}{song} % Use default fonts from CJK (see below)


%\title{Relativistic Brueckner-Hartree-Fock Theory within the continuous choice in the Full Dirac Space}
\title{Partial wave analysis for the in-hadron condensate}

\author{Pianpian Qin}
\affiliation{Department of Physics and Chongqing Key Laboratory for Strongly Coupled Physics, Chongqing University, Chongqing 401331, China}

\author{Zhan Bai}
\affiliation{Insitute of Theoretical Physics,
Chinese Academy of Sciences, Beijing 100190, China}

\author{Muyang Chen}
\affiliation{Department of Physics, Hunan Normal University, Changsha 410081, China}

\author{Si-xue Qin}
\email{sqin@cqu.edu.cn}
\affiliation{Department of Physics and Chongqing Key Laboratory for Strongly Coupled Physics, Chongqing University, Chongqing 401331, China}

\date{\today}

\begin{abstract}
In-hadron condensates, defined as the scalar form factors at zero-momentum transfer, are investigated for flavor-symmetric mesons in pseudoscalar and vector channels under the rainbow-ladder truncation within the Dyson-Schwinger equations framework. 
We confirm the efficiency of the in-hadron condensates in describing the effects of dynamical chiral symmetry breaking from both global and structural perspectives by comparing the meson masses, the dimensionless in-hadron condensates, and the partial wave decompositions of in-hadron condensates as functions of current-quark mass. 
From partial wave analysis, we infer $\pi(1300)$ is a radial excitation dominated by $s$ waves and $\rho(1450)$ is not a $p$ wave-dominated excitation. 
This work provides a new insight into the studies of hadron properties with partial wave analysis for the in-hadron condensates. 
\end{abstract}

%\pacs{26.60.-c,21.65.Ef,21.60.Jz,21.60.De}

%21.60.Jz Nuclear Density Functional Theory and extensions
%21.10.Gv Nucleon distributions and halo features
%21.30.-x Nuclear forces
%21.10.-k Properties of nuclei; nuclear energy levels
%21.10.Re Collective levels
%21.60.Ev Collective models
%21.60.Cs Shell model
%21.45.Ff Three-nucleon forces
%23.20.-g Electromagnetic transitions
%23.20.Js Multipole matrix element
%26.60.-c Nuclear matter aspects of neutron stars
%21.65.Ef Symmetry energy
%21.60.-n Nuclear structure models and methods
%27.20.+n  6 A 19
%27.60.+j 90  A 149
%27.50.+e 59  A  89
%21.60.De Ab initio methods
%\keywords{}

\maketitle


%=======================================================================================
\section{Introduction}\label{SecI}
%=======================================================================================
%begin with the DCSB to vacumm condensate and the in hadron condensate
The chiral symmetry and its breaking are essential in quantum chromo-dynamics (QCD) and play an important role in hadron physics~\cite{Chang2007PRC,Mitter2015PRD,Braun2016PRD,Aguilar2011PRD}. 
Due to the non-vanishing current-quark mass, the chiral symmetry is broken explicitly. 
The current mass of $u/d$ quark is $3$-$5$~MeV. However, the constitute mass of $u/d$ quark inside the nucleon is $300$-$500$~MeV. 
It is believed that this large mass gap stems from the so-called dynamical chiral symmetry breaking~\cite{Roberts2020Sym,Roberts2021PPNP}. 

%The approaches and why our choice
Theoretical studies on dynamical chiral symmetry breaking require non-perturbative tools, such as effective field models~\cite{Buballa2003PR,Ratti2005PRD,Song2020PRC}, functional renormalization group method~\cite{Pawlowski2005AP,Schaefer2006PPN}, Dyson-Schwinger equations (DSEs) approach~\cite{Bashir2012CTP,Roberts2000PPNP,Alkofers2001PR,Maris2003IJMPE,Fischer2006JPG}, and lattice QCD~\cite{montvay1997quantum,Ding2019PRL}. 
In particular, starting from the first principle, DSEs approach preserves the dynamical chiral symmetry breaking and the quark confinement of QCD simultaneously. 
The DSEs approach has been successfully applied to study the hadron properties~\cite{Roberts1994PPNP,Kekez1996PLB,Aguilar2019EPJA,Barabanov2020PPNP}, the hot and dense nuclear matter~\cite{Fischer2018PPNP,Gomes2021PRD,Isserstedt2021PRD}, and the QCD phase transitions~\cite{Qin2010PRL,Fischer2011PLB,Gao2016PRD}. 

With the DSEs approach, the effects of dynamical chiral symmetry breaking can be reflected in observable quantities such as quark mass function~\cite{Miransky1985PLB,Fischer2003PRD}, sigma-term~\cite{Hannah1999PRD,Flambaum2006FBS}, and hadron decay constants~\cite{Maris1998PLB}. 
The vacuum quark condensate is the order parameter of dynamical chiral symmetry breaking, which can measure the effects more directly~\cite{Maris1997PRC,Zong2003PRD}. 
However, the vacuum quark condensate is not well defined since it badly diverges at finite current-quark mass~\cite{Zong2003PRD,Chen2021PRD}. 
Besides, quarks are confined inside hadrons. 
The confinement effect is not considered in vacuum quark condensate, which might be related to the large cosmological constant~\cite{Brodsky2010PRC,Brodsky2012PRC}. 

In Refs.~\cite{Maris1997PRC,Maris1998PLB}, the \textit{in-hadron condensate} is first proposed from the Gell-Mann-Oakes-Renner (GMOR) relation~\cite{Gell-Mann:1968hlm}. 
This in-hadron order parameter of dynamical chiral symmetry breaking is finite at non-vanishing current-quark mass and might reduce the cosmological constant to the observed value~\cite{Brodsky2009PNASU}. 
However, this original definition is only valid in the pseudoscalar channel and is hard to be extended to other channels. 
In Ref.~\cite{Chang2012PRC}, the scalar form factor at zero-momentum transfer is proved as an equivalent quantity to measure the effects of dynamical chiral symmetry breaking, which can be easily extended to all hadrons~\cite{Flambaum2006FBS}. 
Besides, it can also describe the responses of hadron masses to the current-quark mass, which can help to place constraints on the fundamental constants of nature by using observational data~\cite{Webb1999PRL}. 

The scalar form factor at zero-momentum transfer, named in-hadron condensate hereafter, can not only measure the global effects of dynamical chiral symmetry breaking, but also provide a structural perspective for the effects of dynamical chiral symmetry breaking. 
It is known that hadrons have rich and complicated Dirac tensor structures constrained by Poincar$\acute{e}$ covariance~\cite{Hilger2015EPJC,Bhagwat2007EPJA,Xing2021arXiv}. 
By organizing the Dirac tensors with respect to spin and orbital angular momentum of quarks, \textit{partial wave analysis} can be used to distinguish the contributions of corresponding partial waves to the in-hadron condensates~\cite{Macfarlane1962RMP,Eichmann2016PPNP}. 

In this work, we solve the gap equation for quark propagator, the homogeneous Bethe-Salpeter (BS) equation for BS amplitude, and the inhomogeneous BS equation for scalar vertex within the DSEs framework under the rainbow-ladder (RL) truncation scheme. 
To measure the global effects of dynamical chiral symmetry breaking, the hadron masses and the in-hadron condensates are calculated for the ground and first-excited states of flavor-symmetric mesons in pseudoscalar and vector channels. 
Afterwards, the partial wave analyses are performed to the in-hadron condensates to provide a structural perspective for the effects of dynamical chiral symmetry breaking.

This paper is organized as follows. In Sec.~\ref{Sec:Condensate} and~\ref{Sec:DSE} , the definition of in-hadron condensate and the theoretical framework of DSEs approach are introduced respectively. In Sec.~\ref{Sec:Result}, the calculated results and discussions are presented. Finally, a summary is given in Sec.~\ref{Sec:Summary}. 

%=======================================================================================
\section{ Quark Condensates: from Vacuum to in-Hadron}\label{Sec:Condensate}
%=======================================================================================
In the chiral limit, the vacuum quark condensate is defined as 
\begin{equation}\label{eqn:vacuum_condensate}
-\langle\bar{q}q\rangle^{\textrm{vac}}_{0}=Z_4~\text{tr}\int_q^{\Lambda} S_{\hat{m}=0}(q)\,,
\end{equation}
%%-\langle\bar{q}q\rangle^{\textrm{vac}}_{0}=Z_4\textrm{Tr}[S_{\hat{m}=0}(q)]\,,
where $S_{\hat{m}=0}(q)$ is the quark propagator in the chiral limit with $q$ the momentum and $Z_4$ is the corresponding renormalization constant. 
$\int_q^{\Lambda}$ represents a Poincar$\acute{e}$ invariant regularization of the four-dimensional integral with $\Lambda$ the regularization mass-scale. 
The trace is over color and Dirac space. 

Many successes have been achieved in describing the effects of dynamical chiral symmetry breaking by using the vacuum quark condensate. Nevertheless, it suffers from some problems. 
For example, the vacuum quark condensate is badly divergent for finite current-quark mass. 
Although this divergence can be partly eliminated by using subtraction schemes, ambiguities still exist especially at large current-quark mass~\cite{Chang2007PRC,Gao2016PRD,Chen2021PRD,Williams2006PLB,Braun2020PRD,Bazavov2011PRD,Gao2021PRD}. 
Besides, the cosmological constant derived from the vacuum quark condensate is about $10^{46}$ times larger than the observed value~\cite{Brodsky2009PNASU}. 
It is argued that the quark condensate can be modified by the strong interaction inside the hadrons, where the vacuum quark condensate fails to provide a reasonable and realistic measure~\cite{Brodsky2012PRC}. 

To settle these problems, it is natural to generalize the definition of quark condensate from vacuum to in-hadron. 
As pion is the lightest hadron and the Goldstone boson associated with dynamical chiral symmetry breaking, the in-pion condensate is first investigated. 
In Ref.~\cite{Maris1998PLB}, the Gell-Mann-Oakes-Renner (GMOR) relation~\cite{Gell-Mann:1968hlm} is reconsidered, which connects the pion decay constant $f_{\pi}$ and mass $M_{\pi}$ with the vacuum quark condensate 
\begin{equation}\label{eqn:GMOR_chirallimit}
f_{\pi}^2 M_{\pi}^2 \approx -(m_u+m_d) \left\langle \bar{q}q \right\rangle^{\textrm{vac}}_{0}\,, %+\mathcal{O}(\hat{m}^2)\,,
\end{equation}
where $m_{u/d}$ is the current-quark mass. 
$f_{\pi}$ is related to $\langle 0|\bar{q}\gamma_5\gamma_{\mu} q|\Pi\rangle$ with $\left.\left.\right|\Pi \right\rangle$ the pion state. 
Alternatively, the quantity in the left hand side of Eq.~\eqref{eqn:GMOR_chirallimit} can be obtained with the decay constant $\rho_\pi$ by equating pole terms of the corresponding axial and pseudoscalar vertices in axial-vector Ward Takahashi identity (AV-WTI)~\cite{Maris1998PLB}, 
\begin{equation}\label{eqn:GMOR_pion}
f_{\pi}^{2}M_{\pi}^2=(m_u+m_d)f_\pi\rho_\pi\,,
\end{equation}
where $\rho_\pi$ is defined as $i\rho_\pi = -\langle 0|\bar{q}i\gamma_5 q|\Pi\rangle$. 

Comparing Eqs.~\eqref{eqn:GMOR_chirallimit} and~\eqref{eqn:GMOR_pion}, one can naturally define the in-pion condensate~\cite{Maris1997PRC,Brodsky2010PRC,Chang2012PRC} 
\begin{equation}\label{eqn:in-pion condensate}
-\left\langle \bar{q}q \right\rangle_{\pi}:=f_{\pi}\rho_{\pi}\,.
\end{equation}
This definition is valid for non-vanishing current-quark mass and can return to the vacuum quark condensate in the chiral limit as long as the chiral limit residue of the bound state pole in the pseudoscalar vertex is defined as~\cite{Maris1998PLB} 
\begin{equation}\label{eqn:GT_relation}
\rho_{\pi}^{0}=-\frac{1}{f_{\pi}^0}\left\langle \bar{q}q \right\rangle^{\textrm{vac}}_{0}\,,
\end{equation}
where $f_{\pi}^0$ is the value of $f_{\pi}$ in the chiral limit. 
The in-pion condensate has been used to measure the effects of dynamical chiral symmetry breaking inside the pion~\cite{Maris1997PRC,Maris1998PLB,Brodsky2010PRC}. 
This definition in Eq.~\eqref{eqn:in-pion condensate} is valid only for pseudoscalar mesons and is hard to be extended to other hadrons~\cite{Maris1997PRC}. 

In Ref.~\cite{Chang2012PRC}, it is proved that the in-pion condensate can also be represented through the scalar form factor at zero-momentum transfer $Q^2=0$ 
\begin{equation}
S_{\pi}(0)=-\left\langle \Pi\left| \frac{1}{2}(\bar{u}u+\bar{d}d)\right|\Pi \right\rangle \,.
\end{equation}
Besides, the scalar form factor at zero-momentum transfer can be extended to other channels to measure the effects of dynamical chiral symmetry breaking inside the corresponding hadrons. 
Generally, the scalar form factor at zero-momentum transfer for the flavor-symmetric meson can be represented with the expectation value of the operator $\bar{q}q$ in the meson state $\left.\left.\right|P \right\rangle$~\cite{Chang2012PRC} 
\begin{equation}\label{eqn:Sp1}
	S_{P}(0)=-\left\langle P\left| \bar{q}q\right|P \right\rangle \,,
\end{equation}
where $P$ is the meson momentum and is related to the meson mass $M_P$ via the on-shell condition $P^2=-M_P^2$. 
Constrained by the AV-WTI, the scalar form factor is related to the hadron mass $M_P$ and the current-quark mass $m_q$ by~\cite{Flambaum2006FBS,Ren2015PRD} 
\begin{equation}\label{eqn:mass_formula}
S_P(0)=M_P\frac{\partial M_P}{\partial m_q}\,.
\end{equation}
This relation can be viewed as a consequence of the Feynman-Hellmann theorem~\cite{Feynman1939PR}. 
It is found that the dimensionless quantity $S_P(0)/M_P$ is equivalent to the response of hadron mass to the current-quark mass, therefore it is a measure of the global effects of dynamical chiral symmetry breaking. 

\begin{figure}[htbp]
	 	\centering
	 	\includegraphics[width=8.0cm]{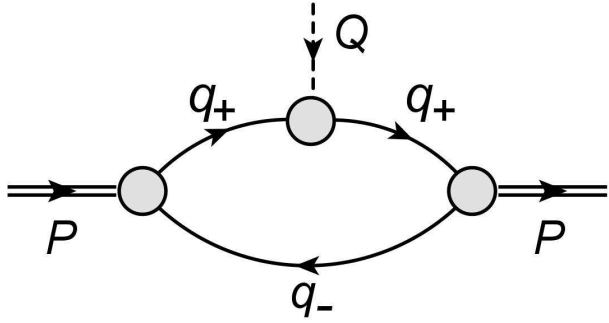}
	 	\caption{The Feynman diagram for the in-hadron condensate in Eq.~\eqref{eqn:Sp2}. $P$ and $q_{\pm}$ are the momenta of the mesons and quark propagators. $Q$ is the transfer momentum with $Q^2=0$.}
	 	\label{fig:fey1-Sp}
 \end{figure}

Within the DSEs framework, the quantity $S_{P}(0)$ can also be calculated from the loop integration 
\begin{equation}\label{eqn:Sp2}
S_P(0)= \text{tr}\int_q^{\Lambda}\bar{\Gamma}_P(q;-P)S(q_+)\Gamma_s(q_+;0)S(q_+)\Gamma_P(q;P)S(q_-)\,,
\end{equation}
where $S(q_{\pm})$ are the renormalized dressed quark propagators with momentum $q_{\pm}=q\pm\frac{1}{2}P$. 
$\Gamma_P(q;P)$ represents the BS amplitude of the meson and $\Gamma_s(q_+;0)$ is the scalar vertex with zero momentum. 
The corresponding Feynman diagram is shown in Fig.~\ref{fig:fey1-Sp}. 
It is obvious that Eq.~\eqref{eqn:Sp2} can be used to analyze the partial wave contributions of the BS amplitudes to the in-hadron condensates, which makes it feasible to probe the effects of dynamical chiral symmetry breaking from a structural perspective. 
 
%fey 1:

\section{Dyson-Schwinger Equation Approach}\label{Sec:DSE}
The equations of motion for Green functions of QCD fields are described by a set of infinite coupled equations, i.e., DSEs. 
The accurate solutions of the DSEs are impossible to obtain. 
To solve the DSEs in practice, the truncation schemes as well as the interaction models are necessary. 
In this work, the RL truncation in Ref.~\cite{Munczek1995PRD,Bender1996PLB} and a simplified version of the interaction in Ref.~\cite{Qin2011PRC} are employed to solve the quark propagator $S(q_{\pm})$, the BS amplitude $\Gamma_P(q;P)$, and the scalar vertex $\Gamma_s(q_+;0)$ in Eq.~\eqref{eqn:Sp2}. 

\subsection{The quark propagator}\label{Sec:GapEquation}
\begin{figure}[htbp]
	 	\centering
	 	\includegraphics[width=12.0cm]{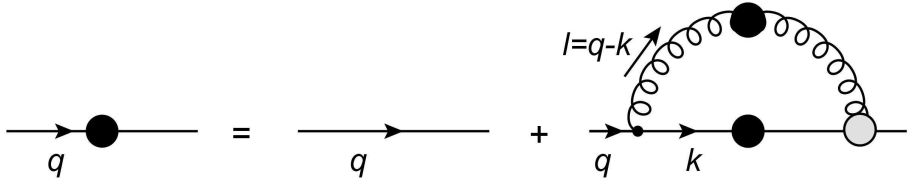}
	 	\caption{The Feynman diagram for the gap equation in Eqs.~\eqref{eqn:DSE} and~\eqref{eqn:self-energy}.}
	 	\label{fig:fey2-DSE}
 \end{figure}

The Dyson-Schwinger equation for quark propagator is known as the gap equation, which reads~\cite{Roberts1994PPNP}
\begin{equation}\label{eqn:DSE}
	S^{-1}(q)=Z_2(i\slashed{q}+Z_m m_q)+\Sigma(q)\,,
\end{equation}
where $\Sigma(q)$ is the self-energy
\begin{equation}\label{eqn:self-energy}
	\Sigma(q)= g^2 Z_1\int_k^{\Lambda} D_{\mu\nu}(l)  \frac{\lambda^a}{2}\gamma_{\mu}S(k) \frac{\lambda^a}{2}\Gamma_{\nu}(k,q)\,.
\end{equation}
The renormalization constants $Z_1$, $Z_2$, and $Z_m$ correspond to the quark-gluon vertex, quark propagator, and current-quark mass, respectively. 
$\frac{\lambda^a}{2}$ is the fundamental representation of $SU(3)$ color symmetry. $D_{\mu\nu}(l)$ with $l=q-k$ is the renormalized dressed gluon propagator and $\Gamma_{\nu}(k,q)$ the renormalized dressed quark-gluon vertex. 
$g$ is the coupling constant. 
The corresponding Feynman diagram of Eqs.~\eqref{eqn:DSE} and~\eqref{eqn:self-energy} is shown in Fig.~\ref{fig:fey2-DSE}. 

%fey 2:
 %

% The quark propagator $S(q)$ can be generally decomposed from tensor analysis as
% %
% \begin{equation}
% 	S(q)=1/[i \slashed{q} A(q^2) +B(q^2)]=Z(q^2)/[i \slashed{q}+M(q^2)]\,,
% \end{equation}
% %
% where $A(q^2)$, $B(q^2)$, and $Z(q^2)$ are scalar functions and $M(q^2)=B(q^2)/A(q^2)$ is the quark mass function. 

%interaction model
For the dressed gluon propagator and the dressed quark-gluon vertex, the following {\it Ans$\ddot{a}$tz} in Ref.~\cite{Qin2011PRC} is used 
\begin{equation}\label{eqn:Interaction}
	Z_1 g^2 D_{\mu\nu}(l)\Gamma_{\nu}(k,q)=l^2\mathcal{G}(l^2)D_{\mu\nu}^{\text{free}}(l)\gamma_{\nu} \,,
\end{equation}
where the dressed quark-gluon vertex $\Gamma_{\nu}(k,q)$ in the quark self-energy in Eq.~\eqref{eqn:self-energy} is truncated to the tree level $\gamma_{\nu}$. 
$D_{\mu\nu}^{\text{free}}(l)=\left(\delta_{\mu\nu}-l_{\mu}l_{\nu}/l^2\right)/l^2$ is the free gluon propagator in Landau gauge. 
The non-perturbative dressing effect is absorbed in the effective interaction function $\mathcal{G}(l^2)$, which is supposed to compensate for all the missing pieces in the quark-gluon vertex~\cite{Qin2011PRC,Maris1999PRC}. 
As the non-perturbative features are insensitive to the ultraviolet behavior of the interaction, in this work we adopt a simplified version of the interaction in Ref.~\cite{Qin2011PRC}, where only the infrared part is kept 
\begin{equation}\label{eqn:QC}
	\mathcal{G}(l^2)=\frac{8\pi^2}{\omega^4}D e^{-l^2/\omega^2}\,.
\end{equation}
In this case the renormalization procedure can be skipped. 
In Eq.~\eqref{eqn:QC} the parameters $D$ and $\omega$ control the strength and width of the interaction respectively. 
It is found that the observables of vector and pseudoscalar mesons are insensitive to variations of $\omega\in[0.4,0.6]$ as long as $D\omega=\text{const.}$ 
In this work, $\omega=0.5$~GeV and $f_{\pi}=0.093$~GeV is obtained with $D\omega=(0.85~\text{GeV})^3$.

\subsection{The Bethe-Salpeter amplitude of the mesons}
\begin{figure}[htbp]
	 	\centering
	 	\includegraphics[width=10.0cm]{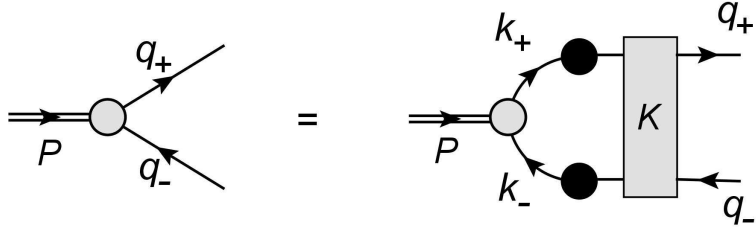}
	 	\caption{The Feynman diagram for the homogeneous BS equation in Eq.~\eqref{eqn:HBSE}.}
	 	\label{fig:fey3-hBSE}
 \end{figure}

%BSE
The BS amplitude of the mesons can be solved from the renormalized homogeneous BS equation, which reads 
\begin{equation}\label{eqn:HBSE}
[\Gamma_P(q;P)]_{\alpha\beta}=\int_{k}^{\Lambda}K_{\alpha\gamma,\beta\delta}(q,k;P)\left[S(k_+)\Gamma_P(k;P)S(k_-)\right]_{\gamma\delta}\,.
\end{equation}
$K_{\alpha\gamma,\delta\beta}(q,k;P)$ is the quark-antiquark scattering kernel with $\alpha$, $\beta$, $\gamma$, $\delta$ the Dirac and color indices. 
For this eigen equation, solutions exist only for particular, separated values of $P^2$. The corresponding Feynman diagram is shown in Fig.~\ref{fig:fey3-hBSE}.

%fey 3:

To solve the homogeneous BS equation, the truncation of the quark-antiquark scattering kernel $K$ is performed with respect to the constraints of AV-WTI~\cite{Chang2021PRD,Qin2021CPL,Qp2019PRD}. 
Considering that the quark-gluon vertex is truncated to the tree level in Eq.~\eqref{eqn:Interaction}, here we choose the one-gluon exchange for the scattering kernel
\begin{equation}\label{eqn:Kernel-Ladder}
K_{\alpha\gamma,\beta\delta}(q,k;P)=-\mathcal{G}(l^2)l^2D_{\mu\nu}^{\text{free}}(l)
\left(\gamma_{\mu}\frac{\lambda^a}{2}\right)_{\alpha\gamma}
\left(\gamma_{\nu}\frac{\lambda^a}{2}\right)_{\beta\delta}\,.
\end{equation}
The united truncation scheme combining Eq.~\eqref{eqn:Interaction} for the quark-gluon vertex and Eq.~\eqref{eqn:Kernel-Ladder} for the scattering kernel is the so-called RL truncation~\cite{Cloet2014PPNP}. It is the first term in a non-perturbative, systematic, and symmetry-preserving approximation. 
RL truncation preserves the one-loop renormalization group properties of QCD, and has provided a uniformly accurate description and prediction for a wide range of hadron properties~\cite{Chang2011CJP,Bashir2012CTP,Qin2019FBS,Aguilar2009PRD}. 

For mesons with spin-parity $J^P$, the BS amplitude $\Gamma_P(q;P)$ can be expanded in the corresponding tensor basis $\tau_P^i(q;P)$ 
\begin{equation}\label{eqn:BSA-expansion}
	\Gamma_P(q;P)=\sum_{i=1}^{N}\tau_P^i(q;P)\mathcal{F}_i(q^2,q\cdot P)\,,
\end{equation}	
where $\mathcal{F}_i(q^2,q\cdot P)$ is the scalar coefficient function. 
$N$ is the dimension of the tensor basis determined by the meson spin $J$. 
In this work, we focus on the pseudoscalar ($J^P=0^-$) and vector ($J^P=1^-$) mesons at ground states and first-excited states, where $N=4$ for $J=0$ and $N=8$ for $J=1$. 
The orthogonal Dirac bases are summarized in Tab.~\ref{tab:PSbasis} and Tab.~\ref{tab:VCbasis} for pseudoscalar and vector mesons respectively~\cite{Hilger2015EPJC}. 
The basis elements are organized with respect to their quark spin $S$ and orbital angular momentum $L$ in the meson's rest frame, where $L=0, 1, 2$ correspond to $s$, $p$, $d$ waves respectively~\cite{Eichmann2016PPNP}. 
In the Dirac bases, $\hat{P}=P/|P|$ is the unit vector of the meson momentum, $q^{\mu}_{\perp} = q^{\mu} - q\cdot\hat{P}\hat{P}^{\mu}$ is the relative momentum transverse to the corresponding meson momentum $P$. 
The Dirac matrix $\gamma^{\mu}_{\perp}=\gamma^{\mu}-\gamma\cdot\hat{P}\hat{P}^{\mu}$ is also transverse to $P$. 

\begin{table}[h]
	\centering
	\caption{The orthogonal Dirac basis for pseudoscalar ($J^P=0^-$) channel with $S$ and $L$ the spin and the orbital angular momentum.}
	\label{tab:PSbasis}
	\begin{tabular}{clcc}
		i & $\tau^i$                                           & S & L  \\ \hline
		1 & $\gamma_{5}$                                    & 0 & 0   \\
		2 & $\gamma_{5}\gamma\cdot\hat{P}$                  & 0 & 0   \\
		3 & $\gamma_{5}\gamma\cdot q_{\perp}$               & 1 & 1   \\
		4 & $\gamma_{5}[\gamma\cdot q,\gamma\cdot\hat{P}]$  & 1 & 1   \\
	\end{tabular}
\end{table}
\begin{table}[h]
	\centering
	\caption{Similar as Tab.~\ref{tab:PSbasis} but for vector ($J^P=1^-$) channel.}
	\label{tab:VCbasis}
	\begin{tabular}{clcc}
		i & $\tau^i$                                                & S & L  \\ \hline
		1 & $\gamma^{\mu}_{\perp}$                               & 1 & 0  \\
		2 & $\gamma^{\mu}_{\perp}\gamma\cdot\hat{P}$             & 1 & 0  \\
		3 & $q^{\mu}_{\perp}\mathbb{1}$                          & 0 & 1  \\
		4 & $q^{\mu}_{\perp}\gamma\cdot\hat{P}$                  & 0 & 1  \\
		5 & $[\gamma^{\mu}_{\perp},\gamma\cdot q_{\perp}]$       & 1 & 1  \\
		6 & $\gamma^{\mu}_{\perp}[\gamma\cdot q,\gamma\cdot\hat{P}]-2q^{\mu}_{\perp}\gamma\cdot\hat{P}$& 1 & 1  \\
		7 & $q^{\mu}_{\perp}\gamma\cdot q_{\perp}-\frac{1}{3}q_{\perp}^2\gamma^{\mu}_{\perp}$    & 1 & 2  \\
		8 & $q^{\mu}_{\perp}[\gamma\cdot q,\gamma\cdot \hat{P}] -\frac{1}{3}q^2_{\perp}[\gamma^{\mu}_{\perp},\gamma\cdot\hat{P}]$ & 1 & 2  \\
	\end{tabular}
\end{table}

\subsection{The scalar vertex}
\begin{figure}[htbp]
	 	\centering
	 	\includegraphics[width=12.0cm]{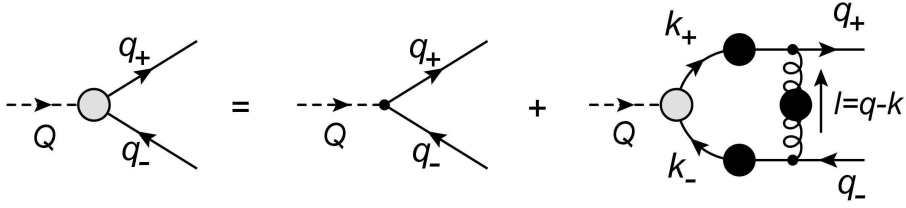}
	 	\caption{The Feynman diagram for the scalar vertex in Eq~\eqref{eqn:iBSE}.}
	 	\label{fig:fey4-iBSE}
 \end{figure}

%vertex
The scalar vertex can be obtained by solving the in-homogeneous BS equation. With RL truncation, it reads 
\begin{equation}\label{eqn:iBSE}
	\Gamma_s(q;Q) = \mathbb{1} +g^2\int_k^{\Lambda}D_{\mu\nu}(l)\frac{\lambda^a}{2}\gamma_{\mu}S(k_+)\Gamma_s(k;Q)S(k_-)\frac{\lambda^a}{2}\gamma_{\nu}\,,
\end{equation}
where $D_{\mu\nu}$ is the dressed gluon propagator with Ans$\ddot{a}$tz given in Eq.~\eqref{eqn:Interaction}. 
$S(k_{\pm})$ are the quark propagator as the solutions of Eqs.~\eqref{eqn:DSE} and~\eqref{eqn:self-energy}. 
The Feynman diagram is shown in Fig.~\ref{fig:fey4-iBSE}.

%fey 4:

Eq.~\eqref{eqn:iBSE} can be solved with iterative algorithm and the solution can be expanded with the full Dirac basis in scalar channel, 
\begin{equation}
	 \{\mathbb{1}\,, \gamma\cdot Q\,, \gamma \cdot q\,, [\slashed{q},\slashed{Q}]\}\,.
\end{equation}
Since the in-hadron condensate is defined as the scalar form factor at zero-momentum transfer $Q^2=0$ in Eqs.~\eqref{eqn:Sp1} and~\eqref{eqn:Sp2}, the Dirac basis is reduced to $\{\mathbb{1}\,,  \gamma \cdot q\}$. 

%=======================================================================================
\section{Results and discussion}\label{Sec:Result}
%=======================================================================================
%fig 1:
%
\begin{figure}[htbp]
	\centering
	\includegraphics[width=10.0cm]{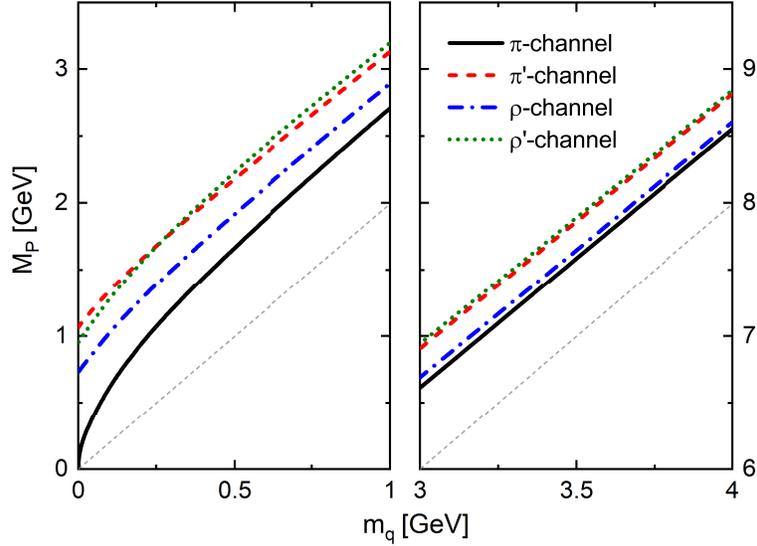}
	\caption{(Color online) The masses of the mesons $M_P$ as functions of the current-quark mass $m_q$ in different channels: pseudoscalar ground state ($\pi$-channel), pseudoscalar first-excited state ($\pi'$-channel), vector ground state ($\rho$-channel) and vector first-excited state ($\rho'$-channel). The grey dashed line represents $2m_q$. The light quark region $[0,1]$~GeV (left) and the heavy quark region $[3,4]$~GeV (right) are shown.}
	\label{fig:fg1-M}
\end{figure}

To measure the total effects of chiral symmetry breaking including both the dynamical and explicit parts, the masses of the pseudoscalar and vector mesons at ground and first-excited states are investigated as shown in Fig.~\ref{fig:fg1-M}.
%Mass-global bahavior
It is obvious that the meson masses $M_P$ increase monotonically with the increase of the current-quark mass $m_q$. 
%%heavy
In the heavy quark region shown in the right panel, similar linear tendencies are found for all the channels. At $m_q\lesssim 4$~GeV, the relative difference between $M_P$ and $2m_q$ are much smaller than that at  $m_q\lesssim 1$~GeV for all these channels, which indicates the effects of dynamical chiral symmetry breaking are weaker compared to that of explicit chiral symmetry breaking for heavy quarks. 
%%light
In the chiral limit shown in the left panel, the mass of the Goldstone boson $\pi$ is vanished. 
With the increase of $m_q$, $M_{\pi}$ increases approximately proportional to $\sqrt{m_q}$ according to Eq.~\eqref{eqn:GMOR_chirallimit}. This nonlinearity indicates the interactions between the quarks inside $\pi$ are significant, which clearly demonstrates the non-perturbative nature of the light quarks. 
Besides, in the chiral limit, the vector meson $\rho$ and the excited states $\pi'$, $\rho'$ possess non-vanishing masses, which are generated totally from the dynamical chiral symmetry breaking. 

%relative size
It is expected that the vector meson is heavier than the pseudoscalar meson for both ground and excited states. 
%%heavy
This is the case for heavy quarks as shown in the right panel in Fig.~\ref{fig:fg1-M}. 
%%light
However, for light quarks near the chiral limit, the mass of the first-excited state in vector channel is found smaller than that in pseudoscalar channel, i.e., $M_{\rho'}<M_{\pi'}$. 
With the increase of the current-quark mass, there exists a crossing point at about $0.3$~GeV, after which the mass ordering of the first-excited states are changed to be $M_{\rho'}>M_{\pi'}$ and remains stable till the heavy quark limit. 

%fig 2: 
%
\begin{figure}[htbp]
	\centering
	\includegraphics[width=10.0cm]{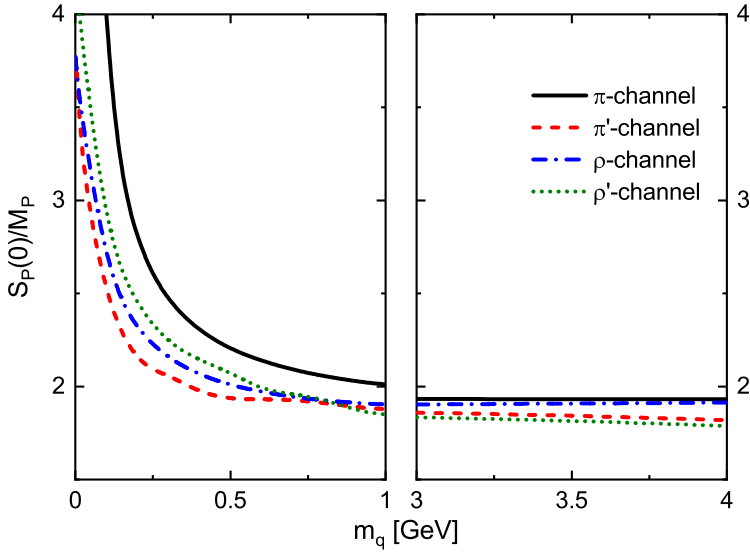}
	\caption{(Color online) Similar as Fig.~\ref{fig:fg1-M} but for dimensionless in-hadron condensates $S_P(0)/M_P$.}
	\label{fig:fg2-S}
\end{figure}
%

%in-hadron condensate-global behavior
To measure the effects of dynamical chiral symmetry breaking, the dimensionless in-hadron condensates $S_P(0)/M_P$ are calculated as functions of current-quark mass $m_q$ for different channels as shown in Fig.~\ref{fig:fg2-S}.
%%heavy
In the heavy quark region, the current-quark mass dependencies of $S_P(0)/M_P$ are quite weak and the differences among different channels are negligible. 
According to Eq.~\eqref{eqn:Sp1}, this behavior of $S_P(0)/M_P$ indicates that the responses of meson masses $M_P$ to the current-quark mass $m_q$ are linear, which is consistent with the results shown in the right panel of Fig.~\ref{fig:fg1-M}. 
%%light
Near the chiral limit, the values of $S_P(0)/M_P$ are significant and decrease rapidly as the current-quark mass increases. 
Since in the chiral limit, the meson masses are generated totally from dynamical chiral symmetry breaking, the rapid decreasing of $S_P(0)/M_P$ probably implies that the global effects of dynamical chiral symmetry breaking is also decreasing. 
%%crossing behavior
Besides, with the increase of the current-quark mass, a crossing behavior for $S_P(0)/M_P$ between the first-excited states in pseudoscalar and vector channels is also found. After the crossing point at about $0.8$~GeV, the ordering of the $S_P(0)/M_P$ remains stable till the heavy quark limit. 
By comparing the meson masses $M_P$ in Fig.~\ref{fig:fg1-M} and the dimensionless in-hadron condensates $S_P(0)/M_P$ in Fig.~\ref{fig:fg2-S}, we find that $S_P(0)/M_P$ and $M_P$ are consistent in measuring the global effects of dynamical chiral symmetry breaking. 

% fig 3: 
%
\begin{figure}[htbp]
  \centering
  \includegraphics[width=10.0cm]{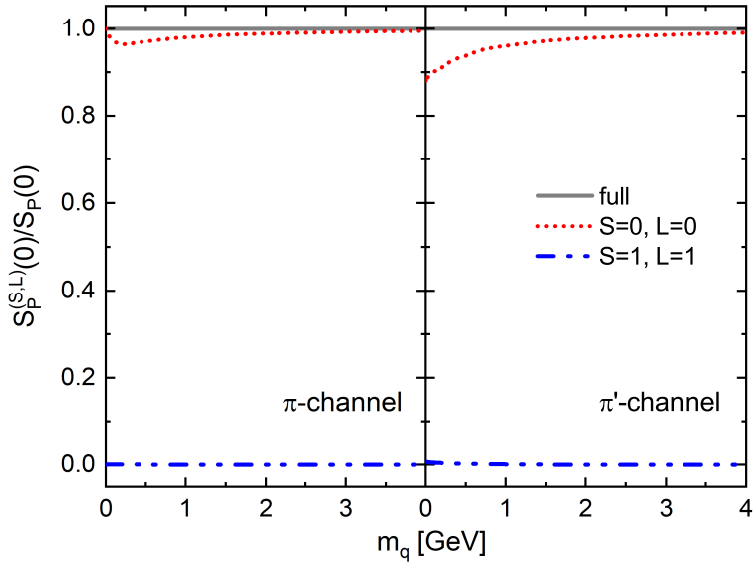}
  \caption{(Color online) The relative partial wave contributions to the in-hadron condensates $S^{(S,L)}_P(0)/S_P(0)$ as functions of current-quark mass $m_q$ for pseudoscalar mesons at ground state (left) and first-excited state (right).}
  \label{fig:fg3-PS}
\end{figure}
%

\begin{figure}[htbp]
	\centering
	\includegraphics[width=10.0cm]{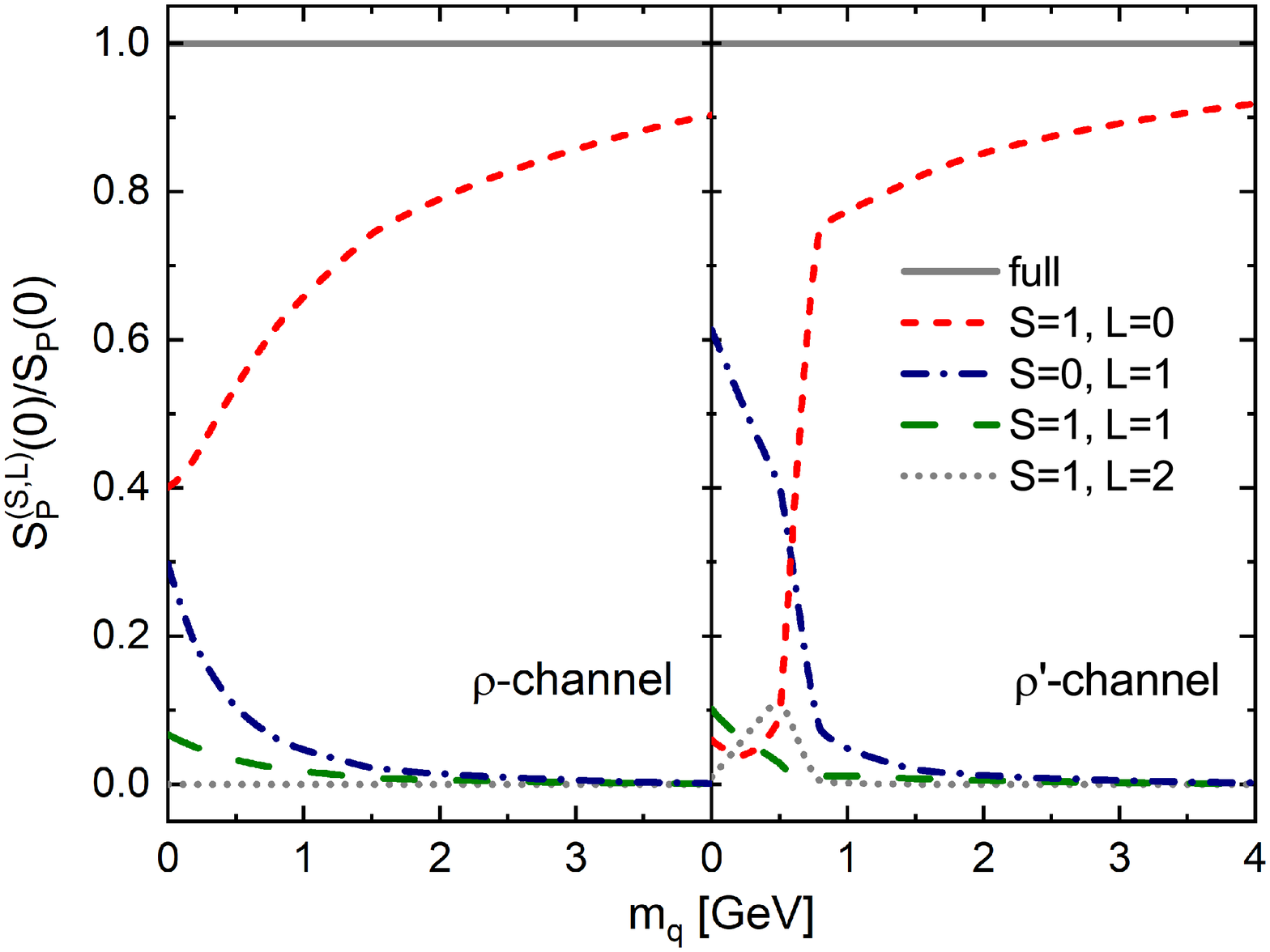}
	\caption{(Color online) Similar as Fig.~\ref{fig:fg3-PS} but for vector mesons.}
	\label{fig:fg4-VC}
\end{figure}

%PartialWaveAnalysis-pseudoscalar
To probe the effects of dynamical chiral symmetry breaking from a structural perspective, the relative partial wave contributions to the in-hadron condensates $S^{(S,L)}_P(0)/S_P(0)$ for pseudoscalar and vector mesons are shown in Fig.~\ref{fig:fg3-PS} and Fig.~\ref{fig:fg4-VC}, respectively. 
The cross terms are not depicted as their contributions can be easily obtained and not of interest here. 
%global
For pseudoscalar mesons at both ground and first-excited states, it is clear that $s$ waves with $(S,L)=(0,0)$ contribute the majority of the in-hadron condensates. 
%%s-wave
This indicates that the dynamical chiral symmetry breaking is mainly contributed by the $s$ waves.
%%pi(1300)
The ordering of the partial wave contributions of the first-excited state is the same as that of the ground state, which indicates the first-excited state in pseudoscalar channel is a radial excitation. 
At the current-quark mass $m_{u/d}=5$~MeV, the mass of the first-excited state in pseudoscalar channel obtained in this work is $1080$~MeV, which is close to the mass of the detected $\pi(1300)$ in experiment. 
This indicates that $\pi(1300)$ is probably a radial excitation dominated by $s$ waves.

%fig 4:
%
%

%PartialWaveAnalysis-vector
%%global
For vector mesons at ground state in the left panel of Fig.~\ref{fig:fg4-VC}, the contribution of the $s$ waves with $(S,L)=(1,0)$ dominates the in-hadron condensates and increases steadily with the increase of the current-quark mass. 
%%p-wave
The second important contribution is from $p$ waves with $(S,L)=(0,1)$, which shows a decreasing tendency with respect to the current-quark mass. 
Similar behavior is found for $p$ waves with $(S,L)=(1,1)$ with less contribution. 
%%d-wave
The contribution of $d$ waves is negligible. 
%%heavy
In the heavy quark limit, the contributions of $p$ and $d$ waves are vanished and the $s$ waves contribute the vast majority. 

%%first-excited
For the first-excited state of vector mesons shown in the right panel of Fig.~\ref{fig:fg4-VC}, the relative partial wave contributions to the in-hadron condensates $S^{(S,L)}_P(0)/S_P(0)$ are found similar as the ground state at heavy quark limit. 
In light quark region, with the increase of the current-quark mass, the contribution of the $s$ waves with $(S,L)=(1,0)$ increase and the contribution of the $p$ waves with $(S,L)=(0,1)$ decrease rapidly. 
A crossing point is found at about $0.7$~GeV. 
Besides, the contributions of $p$ waves with $(S,L)=(1,1)$ and $d$ waves with $(S,L)=(1,2)$ are affected by this novel behavior of the $s$ and $p$ waves. 
Near the chiral limit, the ordering of the partial wave contributions of the first-excited state is different from that of the ground state, which indicates the first-excited state in vector channel is not a radial excitation. 
At the current-quark mass $m_{u/d}=5$~MeV, the mass of the first-excited state in vector channel obtained in this work is $970$~MeV, which is largely different from the mass of the detected $\rho(1450)$ in experiment. 
This indicates that $\rho(1450)$ is not dominated by $p$ waves to a large degree.

%global and structural
From the global perspective in Fig.~\ref{fig:fg2-S}, 
in the heavy quark limit, the ordering of the in-hadron condensates of the ground and first-excited states in pseudoscalar and vector channels is as expected. 
In the light quark region, the crossing point of the in-hadron condensates between the first-excited states in pseudoscalar and vector channels is about $0.8$~GeV. 
From the structural perspective in Fig.~\ref{fig:fg3-PS} and Fig.~\ref{fig:fg4-VC}, in the heavy quark limit, the partial wave contributions to the in-hadron condensates of the ground and first-excited states in pseudoscalar and vector channels are all dominated by $s$ waves. 
In the light quark region, the crossing point of the contributions of $s$ and $p$ waves for the first-excited state in vector channel is about $0.7$~GeV. 
These clearly demonstrate that, on the one hand, it is consistent between the global and the structural perspective of in-hadron condensates in describing the dynamical chiral symmetry breaking. 
On the other hand, the global effects of dynamical chiral symmetry breaking can be understood with the partial wave contributions to in-hadron condensates. 
Furthermore, the crossing point of the meson masses in Fig.~\ref{fig:fg1-M} is smaller than that of the in-hardon condensates in Fig.~\ref{fig:fg2-S}. 
With the increase of current-quark mass, the ordering of the meson masses is as expected after $0.3$~GeV, while the ordering of the in-hardon condensates is not as expected until $0.8$~GeV. That is to say, in the current-quark mass region at about $0.3$-$0.8$~GeV, the meson mass spectra seem reasonable, but are not supported by the structural information of mesons. 
This indicates that mass is not a precise measure to the dynamical chiral symmetry breaking. In comparison, the in-hardron condensate can not only measure the global effects of dynamical chiral symmetry breaking, but also provide a structural perspective for the effects of dynamical chiral symmetry breaking. 

It should be noticed that the meson masses, the in-hadron condensates and the partial wave contributions are all obtained with the DSEs approach under the RL truncation, which truncates the quark-gluon vertex and the quark-antiquark scattering kernel to the tree level and preserves the AV-WTI. 
As the pseudoscalar mesons are strongly constrained by the symmetry embodied in AV-WTI, the calculated properties of the first-excited state in pseudoscalar channel is reliable qualitatively. 
While for the first-excited state of vector meson, the missing contribution from the higher orders might lead to non-negligible effects on spin-orbital interaction and thus on the partial waves, especially when the current-quark mass is close to the chiral limit. 
Thus the relative partial wave contributions to in-hadron condensates might be changed beyond the RL truncation. 

Attempts have been made beyond the RL truncation~\cite{Matevosyan2007PRC,Chang2009PRL,Chang2012PRC85,Binosi2016PRD,Qin2021CPL}. 
In Ref.~\cite{Qin2021CPL}, the dressed-quark anomalous chromomagnetic moment is included and the mass ordering of the first-excited states in pseudoscalar and vector channels has been corrected. 
In future work, it is interesting to study the meson mass spectra and the relative partial wave contributions to in-hadron condensates with a more realistic truncation scheme.

%=======================================================================================
\section{Summary}\label{Sec:Summary}
%=======================================================================================
By using the DSEs approach, the scalar form factors at zero-momentum transfer are calculated as the in-hadron condensates for the ground and first-excited states of the flavor-symmetric mesons in pseudoscalar and vector channels. 
The responses of the dimensionless in-hadron condensates and the meson masses to the current-quark mass are consistent in describing the global effects of dynamical chiral symmetry breaking. 
In-hadron condensate also provides a structural perspective to the effects of dynamical chiral symmetry breaking by comparing the relative partial wave contributions. 
In pseudoscalar channel, for the ground and first-excited states, the in-hardon condensates are dominated by the $s$ wave contributions. 
In vector channel, for the ground state, the in-hardon condensates are dominated by the $s$ wave contributions, while for the first-excited state, they are dominated by the $p$ wave contributions. 
By comparing the theoretical results of the meson masses in pseudoscalar and vector channels with experimental values, we infer that the $\pi(1300)$ is a radial excitation dominated by $s$ waves, while the $\rho(1450)$ is not a $p$ wave-dominated excitation. 

Comparing the crossing behaviors of the meson masses $M_P$, the dimensionless in-hadron condensates $S_P(0)/M_P$, and the partial wave contributions $S^{(S,L)}_P(0)/S_P(0)$, we confirm the efficiency of the in-hadron condensates in describing the effects of dynamical chiral symmetry breaking from both global and structural perspectives. 
This work provides a new insight into the studies of hadron properties with partial wave analysis for the in-hadron condensates. 
In future, we plan to extend this analysis with a more realistic scheme beyond the RL truncation. 

\begin{acknowledgments}
P. Q. thanks Dr. Langtian Liu for helpful discussion. 
This work was partly supported by the National Natural Science Foundation of China under Grant No. 12147102 and No. 12005060, the China Postdoctoral Science Foundation under Grant No. 2021M700610.

\end{acknowledgments}

%\bibliography{ref}

\begin{thebibliography}{68}%
	\makeatletter
	\providecommand \@ifxundefined [1]{%
		\@ifx{#1\undefined}
	}%
	\providecommand \@ifnum [1]{%
		\ifnum #1\expandafter \@firstoftwo
		\else \expandafter \@secondoftwo
		\fi
	}%
	\providecommand \@ifx [1]{%
		\ifx #1\expandafter \@firstoftwo
		\else \expandafter \@secondoftwo
		\fi
	}%
	\providecommand \natexlab [1]{#1}%
	\providecommand \enquote  [1]{``#1''}%
	\providecommand \bibnamefont  [1]{#1}%
	\providecommand \bibfnamefont [1]{#1}%
	\providecommand \citenamefont [1]{#1}%
	\providecommand \href@noop [0]{\@secondoftwo}%
	\providecommand \href [0]{\begingroup \@sanitize@url \@href}%
	\providecommand \@href[1]{\@@startlink{#1}\@@href}%
	\providecommand \@@href[1]{\endgroup#1\@@endlink}%
	\providecommand \@sanitize@url [0]{\catcode `\\12\catcode `\$12\catcode
		`\&12\catcode `\#12\catcode `\^12\catcode `\_12\catcode `\%12\relax}%
	\providecommand \@@startlink[1]{}%
	\providecommand \@@endlink[0]{}%
	\providecommand \url  [0]{\begingroup\@sanitize@url \@url }%
	\providecommand \@url [1]{\endgroup\@href {#1}{\urlprefix }}%
	\providecommand \urlprefix  [0]{URL }%
	\providecommand \Eprint [0]{\href }%
	\providecommand \doibase [0]{http://dx.doi.org/}%
	\providecommand \selectlanguage [0]{\@gobble}%
	\providecommand \bibinfo  [0]{\@secondoftwo}%
	\providecommand \bibfield  [0]{\@secondoftwo}%
	\providecommand \translation [1]{[#1]}%
	\providecommand \BibitemOpen [0]{}%
	\providecommand \bibitemStop [0]{}%
	\providecommand \bibitemNoStop [0]{.\EOS\space}%
	\providecommand \EOS [0]{\spacefactor3000\relax}%
	\providecommand \BibitemShut  [1]{\csname bibitem#1\endcsname}%
	\let\auto@bib@innerbib\@empty
	%</preamble>
	\bibitem [{\citenamefont {Chang}\ \emph {et~al.}(2007)\citenamefont {Chang},
		\citenamefont {Liu}, \citenamefont {Bhagwat}, \citenamefont {Roberts},\ and\
		\citenamefont {Wright}}]{Chang2007PRC}%
	\BibitemOpen
	\bibfield  {author} {\bibinfo {author} {\bibfnamefont {L.}~\bibnamefont
			{Chang}}, \bibinfo {author} {\bibfnamefont {Y.-X.}\ \bibnamefont {Liu}},
		\bibinfo {author} {\bibfnamefont {M.~S.}\ \bibnamefont {Bhagwat}}, \bibinfo
		{author} {\bibfnamefont {C.~D.}\ \bibnamefont {Roberts}}, \ and\ \bibinfo
		{author} {\bibfnamefont {S.~V.}\ \bibnamefont {Wright}},\ }\href {\doibase
		10.1103/PhysRevC.75.015201} {\bibfield  {journal} {\bibinfo  {journal} {Phys.
				Rev. C}\ }\textbf {\bibinfo {volume} {75}},\ \bibinfo {pages} {015201}
		(\bibinfo {year} {2007})}\BibitemShut {NoStop}%
	\bibitem [{\citenamefont {Mitter}\ \emph {et~al.}(2015)\citenamefont {Mitter},
		\citenamefont {Pawlowski},\ and\ \citenamefont {Strodthoff}}]{Mitter2015PRD}%
	\BibitemOpen
	\bibfield  {author} {\bibinfo {author} {\bibfnamefont {M.}~\bibnamefont
			{Mitter}}, \bibinfo {author} {\bibfnamefont {J.~M.}\ \bibnamefont
			{Pawlowski}}, \ and\ \bibinfo {author} {\bibfnamefont {N.}~\bibnamefont
			{Strodthoff}},\ }\href {\doibase 10.1103/PhysRevD.91.054035} {\bibfield
		{journal} {\bibinfo  {journal} {Phys. Rev. D}\ }\textbf {\bibinfo {volume}
			{91}},\ \bibinfo {pages} {054035} (\bibinfo {year} {2015})}\BibitemShut
	{NoStop}%
	\bibitem [{\citenamefont {Braun}\ \emph {et~al.}(2016)\citenamefont {Braun},
		\citenamefont {Fister}, \citenamefont {Pawlowski},\ and\ \citenamefont
		{Rennecke}}]{Braun2016PRD}%
	\BibitemOpen
	\bibfield  {author} {\bibinfo {author} {\bibfnamefont {J.}~\bibnamefont
			{Braun}}, \bibinfo {author} {\bibfnamefont {L.}~\bibnamefont {Fister}},
		\bibinfo {author} {\bibfnamefont {J.~M.}\ \bibnamefont {Pawlowski}}, \ and\
		\bibinfo {author} {\bibfnamefont {F.}~\bibnamefont {Rennecke}},\ }\href
	{\doibase 10.1103/PhysRevD.94.034016} {\bibfield  {journal} {\bibinfo
			{journal} {Phys. Rev. D}\ }\textbf {\bibinfo {volume} {94}},\ \bibinfo
		{pages} {034016} (\bibinfo {year} {2016})}\BibitemShut {NoStop}%
	\bibitem [{\citenamefont {Aguilar}\ and\ \citenamefont
		{Papavassiliou}(2011)}]{Aguilar2011PRD}%
	\BibitemOpen
	\bibfield  {author} {\bibinfo {author} {\bibfnamefont {A.~C.}\ \bibnamefont
			{Aguilar}}\ and\ \bibinfo {author} {\bibfnamefont {J.}~\bibnamefont
			{Papavassiliou}},\ }\href {\doibase 10.1103/PhysRevD.83.014013} {\bibfield
		{journal} {\bibinfo  {journal} {Phys. Rev. D}\ }\textbf {\bibinfo {volume}
			{83}},\ \bibinfo {pages} {014013} (\bibinfo {year} {2011})}\BibitemShut
	{NoStop}%
	\bibitem [{\citenamefont {Roberts}(2020)}]{Roberts2020Sym}%
	\BibitemOpen
	\bibfield  {author} {\bibinfo {author} {\bibfnamefont {C.~D.}\ \bibnamefont
			{Roberts}},\ }\href {\doibase 10.3390/sym12091468} {\bibfield  {journal}
		{\bibinfo  {journal} {Symmetry}\ }\textbf {\bibinfo {volume} {12}} (\bibinfo
		{year} {2020}),\ 10.3390/sym12091468}\BibitemShut {NoStop}%
	\bibitem [{\citenamefont {Roberts}\ \emph {et~al.}(2021)\citenamefont
		{Roberts}, \citenamefont {Richards}, \citenamefont {Horn},\ and\
		\citenamefont {Chang}}]{Roberts2021PPNP}%
	\BibitemOpen
	\bibfield  {author} {\bibinfo {author} {\bibfnamefont {C.~D.}\ \bibnamefont
			{Roberts}}, \bibinfo {author} {\bibfnamefont {D.~G.}\ \bibnamefont
			{Richards}}, \bibinfo {author} {\bibfnamefont {T.}~\bibnamefont {Horn}}, \
		and\ \bibinfo {author} {\bibfnamefont {L.}~\bibnamefont {Chang}},\ }\href
	{\doibase https://doi.org/10.1016/j.ppnp.2021.103883} {\bibfield  {journal}
		{\bibinfo  {journal} {Prog. Part. Nucl. Phys.}\ }\textbf {\bibinfo {volume}
			{120}},\ \bibinfo {pages} {103883} (\bibinfo {year} {2021})}\BibitemShut
	{NoStop}%
	\bibitem [{\citenamefont {Buballa}(2005)}]{Buballa2003PR}%
	\BibitemOpen
	\bibfield  {author} {\bibinfo {author} {\bibfnamefont {M.}~\bibnamefont
			{Buballa}},\ }\href {\doibase 10.1016/j.physrep.2004.11.004} {\bibfield
		{journal} {\bibinfo  {journal} {Phys. Rept.}\ }\textbf {\bibinfo {volume}
			{407}},\ \bibinfo {pages} {205} (\bibinfo {year} {2005})}\BibitemShut
	{NoStop}%
	\bibitem [{\citenamefont {Ratti}\ \emph {et~al.}(2006)\citenamefont {Ratti},
		\citenamefont {Thaler},\ and\ \citenamefont {Weise}}]{Ratti2005PRD}%
	\BibitemOpen
	\bibfield  {author} {\bibinfo {author} {\bibfnamefont {C.}~\bibnamefont
			{Ratti}}, \bibinfo {author} {\bibfnamefont {M.~A.}\ \bibnamefont {Thaler}}, \
		and\ \bibinfo {author} {\bibfnamefont {W.}~\bibnamefont {Weise}},\ }\href
	{\doibase 10.1103/PhysRevD.73.014019} {\bibfield  {journal} {\bibinfo
			{journal} {Phys. Rev. D}\ }\textbf {\bibinfo {volume} {73}},\ \bibinfo
		{pages} {014019} (\bibinfo {year} {2006})}\BibitemShut {NoStop}%
	\bibitem [{\citenamefont {Song}\ \emph {et~al.}(2021)\citenamefont {Song},
		\citenamefont {Tolos}, \citenamefont {Wirth}, \citenamefont {Aichelin},\ and\
		\citenamefont {Bratkovskaya}}]{Song2020PRC}%
	\BibitemOpen
	\bibfield  {author} {\bibinfo {author} {\bibfnamefont {T.}~\bibnamefont
			{Song}}, \bibinfo {author} {\bibfnamefont {L.}~\bibnamefont {Tolos}},
		\bibinfo {author} {\bibfnamefont {J.}~\bibnamefont {Wirth}}, \bibinfo
		{author} {\bibfnamefont {J.}~\bibnamefont {Aichelin}}, \ and\ \bibinfo
		{author} {\bibfnamefont {E.}~\bibnamefont {Bratkovskaya}},\ }\href {\doibase
		10.1103/PhysRevC.103.044901} {\bibfield  {journal} {\bibinfo  {journal}
			{Phys. Rev. C}\ }\textbf {\bibinfo {volume} {103}},\ \bibinfo {pages}
		{044901} (\bibinfo {year} {2021})}\BibitemShut {NoStop}%
	\bibitem [{\citenamefont {Pawlowski}(2007)}]{Pawlowski2005AP}%
	\BibitemOpen
	\bibfield  {author} {\bibinfo {author} {\bibfnamefont {J.~M.}\ \bibnamefont
			{Pawlowski}},\ }\href {\doibase 10.1016/j.aop.2007.01.007} {\bibfield
		{journal} {\bibinfo  {journal} {Annals Phys.}\ }\textbf {\bibinfo {volume}
			{322}},\ \bibinfo {pages} {2831} (\bibinfo {year} {2007})}\BibitemShut
	{NoStop}%
	\bibitem [{\citenamefont {Schaefer}\ and\ \citenamefont
		{Wambach}(2008)}]{Schaefer2006PPN}%
	\BibitemOpen
	\bibfield  {author} {\bibinfo {author} {\bibfnamefont {B.-J.}\ \bibnamefont
			{Schaefer}}\ and\ \bibinfo {author} {\bibfnamefont {J.}~\bibnamefont
			{Wambach}},\ }\href {\doibase 10.1134/S1063779608070083} {\bibfield
		{journal} {\bibinfo  {journal} {Phys. Part. Nucl.}\ }\textbf {\bibinfo
			{volume} {39}},\ \bibinfo {pages} {1025} (\bibinfo {year}
		{2008})}\BibitemShut {NoStop}%
	\bibitem [{\citenamefont {Bashir}\ \emph {et~al.}(2012)\citenamefont {Bashir},
		\citenamefont {Chang}, \citenamefont {Cloet}, \citenamefont {El-Bennich},
		\citenamefont {Liu}, \citenamefont {Roberts},\ and\ \citenamefont
		{Tandy}}]{Bashir2012CTP}%
	\BibitemOpen
	\bibfield  {author} {\bibinfo {author} {\bibfnamefont {A.}~\bibnamefont
			{Bashir}}, \bibinfo {author} {\bibfnamefont {L.}~\bibnamefont {Chang}},
		\bibinfo {author} {\bibfnamefont {I.~C.}\ \bibnamefont {Cloet}}, \bibinfo
		{author} {\bibfnamefont {B.}~\bibnamefont {El-Bennich}}, \bibinfo {author}
		{\bibfnamefont {Y.-X.}\ \bibnamefont {Liu}}, \bibinfo {author} {\bibfnamefont
			{C.~D.}\ \bibnamefont {Roberts}}, \ and\ \bibinfo {author} {\bibfnamefont
			{P.~C.}\ \bibnamefont {Tandy}},\ }\href {\doibase 10.1088/0253-6102/58/1/16}
	{\bibfield  {journal} {\bibinfo  {journal} {Commun. Theor. Phys.}\ }\textbf
		{\bibinfo {volume} {58}},\ \bibinfo {pages} {79} (\bibinfo {year}
		{2012})}\BibitemShut {NoStop}%
	\bibitem [{\citenamefont {Roberts}\ and\ \citenamefont
		{Schmidt}(2000)}]{Roberts2000PPNP}%
	\BibitemOpen
	\bibfield  {author} {\bibinfo {author} {\bibfnamefont {C.~D.}\ \bibnamefont
			{Roberts}}\ and\ \bibinfo {author} {\bibfnamefont {S.~M.}\ \bibnamefont
			{Schmidt}},\ }\href {\doibase 10.1016/S0146-6410(00)90011-5} {\bibfield
		{journal} {\bibinfo  {journal} {Prog. Part. Nucl. Phys.}\ }\textbf {\bibinfo
			{volume} {45}},\ \bibinfo {pages} {S1} (\bibinfo {year} {2000})}\BibitemShut
	{NoStop}%
	\bibitem [{\citenamefont {Alkofer}\ and\ \citenamefont {{von
				Smekal}}(2001)}]{Alkofers2001PR}%
	\BibitemOpen
	\bibfield  {author} {\bibinfo {author} {\bibfnamefont {R.}~\bibnamefont
			{Alkofer}}\ and\ \bibinfo {author} {\bibfnamefont {L.}~\bibnamefont {{von
					Smekal}}},\ }\href {\doibase https://doi.org/10.1016/S0370-1573(01)00010-2}
	{\bibfield  {journal} {\bibinfo  {journal} {Phys. Rept.}\ }\textbf {\bibinfo
			{volume} {353}},\ \bibinfo {pages} {281} (\bibinfo {year}
		{2001})}\BibitemShut {NoStop}%
	\bibitem [{\citenamefont {Maris}\ and\ \citenamefont
		{Roberts}(2003)}]{Maris2003IJMPE}%
	\BibitemOpen
	\bibfield  {author} {\bibinfo {author} {\bibfnamefont {P.}~\bibnamefont
			{Maris}}\ and\ \bibinfo {author} {\bibfnamefont {C.~D.}\ \bibnamefont
			{Roberts}},\ }\href {\doibase 10.1142/S0218301303001326} {\bibfield
		{journal} {\bibinfo  {journal} {Int. J. Mod. Phys. E}\ }\textbf {\bibinfo
			{volume} {12}},\ \bibinfo {pages} {297} (\bibinfo {year} {2003})},\ \Eprint
	{http://arxiv.org/abs/nucl-th/0301049} {arXiv:nucl-th/0301049} \BibitemShut
	{NoStop}%
	\bibitem [{\citenamefont {Fischer}(2006)}]{Fischer2006JPG}%
	\BibitemOpen
	\bibfield  {author} {\bibinfo {author} {\bibfnamefont {C.~S.}\ \bibnamefont
			{Fischer}},\ }\href {\doibase 10.1088/0954-3899/32/8/R02} {\bibfield
		{journal} {\bibinfo  {journal} {J. Phys. G}\ }\textbf {\bibinfo {volume}
			{32}},\ \bibinfo {pages} {R253} (\bibinfo {year} {2006})}\BibitemShut
	{NoStop}%
	\bibitem [{\citenamefont {Montvay}\ and\ \citenamefont
		{M{\"u}nster}(1997)}]{montvay1997quantum}%
	\BibitemOpen
	\bibfield  {author} {\bibinfo {author} {\bibfnamefont {I.}~\bibnamefont
			{Montvay}}\ and\ \bibinfo {author} {\bibfnamefont {G.}~\bibnamefont
			{M{\"u}nster}},\ }\href@noop {} {\emph {\bibinfo {title} {Quantum fields on a
				lattice}}}\ (\bibinfo  {publisher} {Cambridge University Press},\ \bibinfo
	{year} {1997})\BibitemShut {NoStop}%
	\bibitem [{\citenamefont {Ding}\ and\ \citenamefont
		{et~al.}(2019)}]{Ding2019PRL}%
	\BibitemOpen
	\bibfield  {author} {\bibinfo {author} {\bibfnamefont {H.~T.}\ \bibnamefont
			{Ding}}\ and\ \bibinfo {author} {\bibnamefont {et~al.}} (\bibinfo
		{collaboration} {HotQCD}),\ }\href {\doibase 10.1103/PhysRevLett.123.062002}
	{\bibfield  {journal} {\bibinfo  {journal} {Phys. Rev. Lett.}\ }\textbf
		{\bibinfo {volume} {123}},\ \bibinfo {pages} {062002} (\bibinfo {year}
		{2019})}\BibitemShut {NoStop}%
	\bibitem [{\citenamefont {Roberts}\ and\ \citenamefont
		{Williams}(1994)}]{Roberts1994PPNP}%
	\BibitemOpen
	\bibfield  {author} {\bibinfo {author} {\bibfnamefont {C.~D.}\ \bibnamefont
			{Roberts}}\ and\ \bibinfo {author} {\bibfnamefont {A.~G.}\ \bibnamefont
			{Williams}},\ }\href {\doibase https://doi.org/10.1016/0146-6410(94)90049-3}
	{\bibfield  {journal} {\bibinfo  {journal} {Prog. Part. Nucl. Phys.}\
		}\textbf {\bibinfo {volume} {33}},\ \bibinfo {pages} {477} (\bibinfo {year}
		{1994})}\BibitemShut {NoStop}%
	\bibitem [{\citenamefont {Kekez}\ and\ \citenamefont
		{Klabučar}(1996)}]{Kekez1996PLB}%
	\BibitemOpen
	\bibfield  {author} {\bibinfo {author} {\bibfnamefont {D.}~\bibnamefont
			{Kekez}}\ and\ \bibinfo {author} {\bibfnamefont {D.}~\bibnamefont
			{Klabučar}},\ }\href {\doibase https://doi.org/10.1016/0370-2693(96)00990-2}
	{\bibfield  {journal} {\bibinfo  {journal} {Phys. Lett. B}\ }\textbf
		{\bibinfo {volume} {387}},\ \bibinfo {pages} {14} (\bibinfo {year}
		{1996})}\BibitemShut {NoStop}%
	\bibitem [{\citenamefont {Aguilar}\ and\ \citenamefont
		{et~al.}(2019)}]{Aguilar2019EPJA}%
	\BibitemOpen
	\bibfield  {author} {\bibinfo {author} {\bibfnamefont {A.~C.}\ \bibnamefont
			{Aguilar}}\ and\ \bibinfo {author} {\bibnamefont {et~al.}},\ }\href {\doibase
		10.1140/epja/i2019-12885-0} {\bibfield  {journal} {\bibinfo  {journal} {Eur.
				Phys. J. A}\ }\textbf {\bibinfo {volume} {55}},\ \bibinfo {pages} {190}
		(\bibinfo {year} {2019})}\BibitemShut {NoStop}%
	\bibitem [{\citenamefont {Barabanov}\ and\ \citenamefont
		{et~al.}(2021)}]{Barabanov2020PPNP}%
	\BibitemOpen
	\bibfield  {author} {\bibinfo {author} {\bibfnamefont {M.~Y.}\ \bibnamefont
			{Barabanov}}\ and\ \bibinfo {author} {\bibnamefont {et~al.}},\ }\href
	{\doibase 10.1016/j.ppnp.2020.103835} {\bibfield  {journal} {\bibinfo
			{journal} {Prog. Part. Nucl. Phys.}\ }\textbf {\bibinfo {volume} {116}},\
		\bibinfo {pages} {103835} (\bibinfo {year} {2021})}\BibitemShut {NoStop}%
	\bibitem [{\citenamefont {Fischer}(2019)}]{Fischer2018PPNP}%
	\BibitemOpen
	\bibfield  {author} {\bibinfo {author} {\bibfnamefont {C.~S.}\ \bibnamefont
			{Fischer}},\ }\href {\doibase 10.1016/j.ppnp.2019.01.002} {\bibfield
		{journal} {\bibinfo  {journal} {Prog. Part. Nucl. Phys.}\ }\textbf {\bibinfo
			{volume} {105}},\ \bibinfo {pages} {1} (\bibinfo {year} {2019})}\BibitemShut
	{NoStop}%
	\bibitem [{\citenamefont {Gomes}(2021)}]{Gomes2021PRD}%
	\BibitemOpen
	\bibfield  {author} {\bibinfo {author} {\bibfnamefont {Y.~M.~P.}\
			\bibnamefont {Gomes}},\ }\href {\doibase 10.1103/PhysRevD.104.015022}
	{\bibfield  {journal} {\bibinfo  {journal} {Phys. Rev. D}\ }\textbf {\bibinfo
			{volume} {104}},\ \bibinfo {pages} {015022} (\bibinfo {year}
		{2021})}\BibitemShut {NoStop}%
	\bibitem [{\citenamefont {Isserstedt}\ \emph {et~al.}(2021)\citenamefont
		{Isserstedt}, \citenamefont {Fischer},\ and\ \citenamefont
		{Steinert}}]{Isserstedt2021PRD}%
	\BibitemOpen
	\bibfield  {author} {\bibinfo {author} {\bibfnamefont {P.}~\bibnamefont
			{Isserstedt}}, \bibinfo {author} {\bibfnamefont {C.~S.}\ \bibnamefont
			{Fischer}}, \ and\ \bibinfo {author} {\bibfnamefont {T.}~\bibnamefont
			{Steinert}},\ }\href {\doibase 10.1103/PhysRevD.103.054012} {\bibfield
		{journal} {\bibinfo  {journal} {Phys. Rev. D}\ }\textbf {\bibinfo {volume}
			{103}},\ \bibinfo {pages} {054012} (\bibinfo {year} {2021})}\BibitemShut
	{NoStop}%
	\bibitem [{\citenamefont {Qin}\ \emph {et~al.}(2011{\natexlab{a}})\citenamefont
		{Qin}, \citenamefont {Chang}, \citenamefont {Chen}, \citenamefont {Liu},\
		and\ \citenamefont {Roberts}}]{Qin2010PRL}%
	\BibitemOpen
	\bibfield  {author} {\bibinfo {author} {\bibfnamefont {S.-X.}\ \bibnamefont
			{Qin}}, \bibinfo {author} {\bibfnamefont {L.}~\bibnamefont {Chang}}, \bibinfo
		{author} {\bibfnamefont {H.}~\bibnamefont {Chen}}, \bibinfo {author}
		{\bibfnamefont {Y.-X.}\ \bibnamefont {Liu}}, \ and\ \bibinfo {author}
		{\bibfnamefont {C.~D.}\ \bibnamefont {Roberts}},\ }\href {\doibase
		10.1103/PhysRevLett.106.172301} {\bibfield  {journal} {\bibinfo  {journal}
			{Phys. Rev. Lett.}\ }\textbf {\bibinfo {volume} {106}},\ \bibinfo {pages}
		{172301} (\bibinfo {year} {2011}{\natexlab{a}})}\BibitemShut {NoStop}%
	\bibitem [{\citenamefont {Fischer}\ \emph {et~al.}(2011)\citenamefont
		{Fischer}, \citenamefont {Luecker},\ and\ \citenamefont
		{Mueller}}]{Fischer2011PLB}%
	\BibitemOpen
	\bibfield  {author} {\bibinfo {author} {\bibfnamefont {C.~S.}\ \bibnamefont
			{Fischer}}, \bibinfo {author} {\bibfnamefont {J.}~\bibnamefont {Luecker}}, \
		and\ \bibinfo {author} {\bibfnamefont {J.~A.}\ \bibnamefont {Mueller}},\
	}\href {\doibase https://doi.org/10.1016/j.physletb.2011.07.039} {\bibfield
		{journal} {\bibinfo  {journal} {Phys. Lett. B}\ }\textbf {\bibinfo {volume}
			{702}},\ \bibinfo {pages} {438} (\bibinfo {year} {2011})}\BibitemShut
	{NoStop}%
	\bibitem [{\citenamefont {Gao}\ and\ \citenamefont {Liu}(2016)}]{Gao2016PRD}%
	\BibitemOpen
	\bibfield  {author} {\bibinfo {author} {\bibfnamefont {F.}~\bibnamefont
			{Gao}}\ and\ \bibinfo {author} {\bibfnamefont {Y.-x.}\ \bibnamefont {Liu}},\
	}\href {\doibase 10.1103/PhysRevD.94.076009} {\bibfield  {journal} {\bibinfo
			{journal} {Phys. Rev. D}\ }\textbf {\bibinfo {volume} {94}},\ \bibinfo
		{pages} {076009} (\bibinfo {year} {2016})}\BibitemShut {NoStop}%
	\bibitem [{\citenamefont {Miransky}(1985)}]{Miransky1985PLB}%
	\BibitemOpen
	\bibfield  {author} {\bibinfo {author} {\bibfnamefont {V.~A.}\ \bibnamefont
			{Miransky}},\ }\href {\doibase 10.1016/0370-2693(85)91254-7} {\bibfield
		{journal} {\bibinfo  {journal} {Phys. Lett. B}\ }\textbf {\bibinfo {volume}
			{165}},\ \bibinfo {pages} {401} (\bibinfo {year} {1985})}\BibitemShut
	{NoStop}%
	\bibitem [{\citenamefont {Fischer}\ and\ \citenamefont
		{Alkofer}(2003)}]{Fischer2003PRD}%
	\BibitemOpen
	\bibfield  {author} {\bibinfo {author} {\bibfnamefont {C.~S.}\ \bibnamefont
			{Fischer}}\ and\ \bibinfo {author} {\bibfnamefont {R.}~\bibnamefont
			{Alkofer}},\ }\href {\doibase 10.1103/PhysRevD.67.094020} {\bibfield
		{journal} {\bibinfo  {journal} {Phys. Rev. D}\ }\textbf {\bibinfo {volume}
			{67}},\ \bibinfo {pages} {094020} (\bibinfo {year} {2003})}\BibitemShut
	{NoStop}%
	\bibitem [{\citenamefont {Hannah}(1999)}]{Hannah1999PRD}%
	\BibitemOpen
	\bibfield  {author} {\bibinfo {author} {\bibfnamefont {T.}~\bibnamefont
			{Hannah}},\ }\href {\doibase 10.1103/PhysRevD.60.017502} {\bibfield
		{journal} {\bibinfo  {journal} {Phys. Rev. D}\ }\textbf {\bibinfo {volume}
			{60}},\ \bibinfo {pages} {017502} (\bibinfo {year} {1999})}\BibitemShut
	{NoStop}%
	\bibitem [{\citenamefont {Flambaum}\ \emph {et~al.}(2006)\citenamefont
		{Flambaum}, \citenamefont {Hoell}, \citenamefont {Jaikumar}, \citenamefont
		{Roberts},\ and\ \citenamefont {Wright}}]{Flambaum2006FBS}%
	\BibitemOpen
	\bibfield  {author} {\bibinfo {author} {\bibfnamefont {V.~V.}\ \bibnamefont
			{Flambaum}}, \bibinfo {author} {\bibfnamefont {A.}~\bibnamefont {Hoell}},
		\bibinfo {author} {\bibfnamefont {P.}~\bibnamefont {Jaikumar}}, \bibinfo
		{author} {\bibfnamefont {C.~D.}\ \bibnamefont {Roberts}}, \ and\ \bibinfo
		{author} {\bibfnamefont {S.~V.}\ \bibnamefont {Wright}},\ }\href {\doibase
		10.1007/s00601-005-0123-1} {\bibfield  {journal} {\bibinfo  {journal} {Few
				Body Syst.}\ }\textbf {\bibinfo {volume} {38}},\ \bibinfo {pages} {31}
		(\bibinfo {year} {2006})}\BibitemShut {NoStop}%
	\bibitem [{\citenamefont {Maris}\ \emph {et~al.}(1998)\citenamefont {Maris},
		\citenamefont {Roberts},\ and\ \citenamefont {Tandy}}]{Maris1998PLB}%
	\BibitemOpen
	\bibfield  {author} {\bibinfo {author} {\bibfnamefont {P.}~\bibnamefont
			{Maris}}, \bibinfo {author} {\bibfnamefont {C.~D.}\ \bibnamefont {Roberts}},
		\ and\ \bibinfo {author} {\bibfnamefont {P.~C.}\ \bibnamefont {Tandy}},\
	}\href {\doibase 10.1016/S0370-2693(97)01535-9} {\bibfield  {journal}
		{\bibinfo  {journal} {Phys. Lett. B}\ }\textbf {\bibinfo {volume} {420}},\
		\bibinfo {pages} {267} (\bibinfo {year} {1998})}\BibitemShut {NoStop}%
	\bibitem [{\citenamefont {Maris}\ and\ \citenamefont
		{Roberts}(1997)}]{Maris1997PRC}%
	\BibitemOpen
	\bibfield  {author} {\bibinfo {author} {\bibfnamefont {P.}~\bibnamefont
			{Maris}}\ and\ \bibinfo {author} {\bibfnamefont {C.~D.}\ \bibnamefont
			{Roberts}},\ }\href {\doibase 10.1103/PhysRevC.56.3369} {\bibfield  {journal}
		{\bibinfo  {journal} {Phys. Rev. C}\ }\textbf {\bibinfo {volume} {56}},\
		\bibinfo {pages} {3369} (\bibinfo {year} {1997})}\BibitemShut {NoStop}%
	\bibitem [{\citenamefont {Zong}\ \emph {et~al.}(2003)\citenamefont {Zong},
		\citenamefont {Ping}, \citenamefont {Yang}, \citenamefont {L\"u},\ and\
		\citenamefont {Wang}}]{Zong2003PRD}%
	\BibitemOpen
	\bibfield  {author} {\bibinfo {author} {\bibfnamefont {H.-S.}\ \bibnamefont
			{Zong}}, \bibinfo {author} {\bibfnamefont {J.-L.}\ \bibnamefont {Ping}},
		\bibinfo {author} {\bibfnamefont {H.-T.}\ \bibnamefont {Yang}}, \bibinfo
		{author} {\bibfnamefont {X.-F.}\ \bibnamefont {L\"u}}, \ and\ \bibinfo
		{author} {\bibfnamefont {F.}~\bibnamefont {Wang}},\ }\href {\doibase
		10.1103/PhysRevD.67.074004} {\bibfield  {journal} {\bibinfo  {journal} {Phys.
				Rev. D}\ }\textbf {\bibinfo {volume} {67}},\ \bibinfo {pages} {074004}
		(\bibinfo {year} {2003})}\BibitemShut {NoStop}%
	\bibitem [{\citenamefont {Chen}\ \emph {et~al.}(2021)\citenamefont {Chen},
		\citenamefont {Bai}, \citenamefont {Gao},\ and\ \citenamefont
		{Liu}}]{Chen2021PRD}%
	\BibitemOpen
	\bibfield  {author} {\bibinfo {author} {\bibfnamefont {L.-F.}\ \bibnamefont
			{Chen}}, \bibinfo {author} {\bibfnamefont {Z.}~\bibnamefont {Bai}}, \bibinfo
		{author} {\bibfnamefont {F.}~\bibnamefont {Gao}}, \ and\ \bibinfo {author}
		{\bibfnamefont {Y.-X.}\ \bibnamefont {Liu}},\ }\href {\doibase
		10.1103/PhysRevD.104.094041} {\bibfield  {journal} {\bibinfo  {journal}
			{Phys. Rev. D}\ }\textbf {\bibinfo {volume} {104}},\ \bibinfo {pages}
		{094041} (\bibinfo {year} {2021})}\BibitemShut {NoStop}%
	\bibitem [{\citenamefont {Brodsky}\ \emph {et~al.}(2010)\citenamefont
		{Brodsky}, \citenamefont {Roberts}, \citenamefont {Shrock},\ and\
		\citenamefont {Tandy}}]{Brodsky2010PRC}%
	\BibitemOpen
	\bibfield  {author} {\bibinfo {author} {\bibfnamefont {S.~J.}\ \bibnamefont
			{Brodsky}}, \bibinfo {author} {\bibfnamefont {C.~D.}\ \bibnamefont
			{Roberts}}, \bibinfo {author} {\bibfnamefont {R.}~\bibnamefont {Shrock}}, \
		and\ \bibinfo {author} {\bibfnamefont {P.~C.}\ \bibnamefont {Tandy}},\ }\href
	{\doibase 10.1103/PhysRevC.82.022201} {\bibfield  {journal} {\bibinfo
			{journal} {Phys. Rev. C}\ }\textbf {\bibinfo {volume} {82}},\ \bibinfo
		{pages} {022201} (\bibinfo {year} {2010})}\BibitemShut {NoStop}%
	\bibitem [{\citenamefont {Brodsky}\ \emph {et~al.}(2012)\citenamefont
		{Brodsky}, \citenamefont {Roberts}, \citenamefont {Shrock},\ and\
		\citenamefont {Tandy}}]{Brodsky2012PRC}%
	\BibitemOpen
	\bibfield  {author} {\bibinfo {author} {\bibfnamefont {S.~J.}\ \bibnamefont
			{Brodsky}}, \bibinfo {author} {\bibfnamefont {C.~D.}\ \bibnamefont
			{Roberts}}, \bibinfo {author} {\bibfnamefont {R.}~\bibnamefont {Shrock}}, \
		and\ \bibinfo {author} {\bibfnamefont {P.~C.}\ \bibnamefont {Tandy}},\ }\href
	{\doibase 10.1103/PhysRevC.85.065202} {\bibfield  {journal} {\bibinfo
			{journal} {Phys. Rev. C}\ }\textbf {\bibinfo {volume} {85}},\ \bibinfo
		{pages} {065202} (\bibinfo {year} {2012})}\BibitemShut {NoStop}%
	\bibitem [{\citenamefont {{Gell-Mann}}\ \emph {et~al.}(1968)\citenamefont
		{{Gell-Mann}}, \citenamefont {Oakes},\ and\ \citenamefont
		{Renner}}]{Gell-Mann:1968hlm}%
	\BibitemOpen
	\bibfield  {author} {\bibinfo {author} {\bibfnamefont {M.}~\bibnamefont
			{{Gell-Mann}}}, \bibinfo {author} {\bibfnamefont {R.~J.}\ \bibnamefont
			{Oakes}}, \ and\ \bibinfo {author} {\bibfnamefont {B.}~\bibnamefont
			{Renner}},\ }\href {\doibase 10.1103/PhysRev.175.2195} {\bibfield  {journal}
		{\bibinfo  {journal} {Phys. Rev.}\ }\textbf {\bibinfo {volume} {175}},\
		\bibinfo {pages} {2195} (\bibinfo {year} {1968})}\BibitemShut {NoStop}%
	\bibitem [{\citenamefont {Brodsky}\ and\ \citenamefont
		{Shrock}(2011)}]{Brodsky2009PNASU}%
	\BibitemOpen
	\bibfield  {author} {\bibinfo {author} {\bibfnamefont {S.~J.}\ \bibnamefont
			{Brodsky}}\ and\ \bibinfo {author} {\bibfnamefont {R.}~\bibnamefont
			{Shrock}},\ }\href {\doibase 10.1073/pnas.1010113107} {\bibfield  {journal}
		{\bibinfo  {journal} {Proc Natl Acad Sci USA}\ }\textbf {\bibinfo {volume}
			{108}},\ \bibinfo {pages} {45} (\bibinfo {year} {2011})}\BibitemShut
	{NoStop}%
	\bibitem [{\citenamefont {Chang}\ \emph {et~al.}(2012)\citenamefont {Chang},
		\citenamefont {Roberts},\ and\ \citenamefont {Tandy}}]{Chang2012PRC}%
	\BibitemOpen
	\bibfield  {author} {\bibinfo {author} {\bibfnamefont {L.}~\bibnamefont
			{Chang}}, \bibinfo {author} {\bibfnamefont {C.~D.}\ \bibnamefont {Roberts}},
		\ and\ \bibinfo {author} {\bibfnamefont {P.~C.}\ \bibnamefont {Tandy}},\
	}\href {\doibase 10.1103/PhysRevC.85.012201} {\bibfield  {journal} {\bibinfo
			{journal} {Phys. Rev. C}\ }\textbf {\bibinfo {volume} {85}},\ \bibinfo
		{pages} {012201} (\bibinfo {year} {2012})}\BibitemShut {NoStop}%
	\bibitem [{\citenamefont {Webb}\ \emph {et~al.}(1999)\citenamefont {Webb},
		\citenamefont {Flambaum}, \citenamefont {Churchill}, \citenamefont
		{Drinkwater},\ and\ \citenamefont {Barrow}}]{Webb1999PRL}%
	\BibitemOpen
	\bibfield  {author} {\bibinfo {author} {\bibfnamefont {J.~K.}\ \bibnamefont
			{Webb}}, \bibinfo {author} {\bibfnamefont {V.~V.}\ \bibnamefont {Flambaum}},
		\bibinfo {author} {\bibfnamefont {C.~W.}\ \bibnamefont {Churchill}}, \bibinfo
		{author} {\bibfnamefont {M.~J.}\ \bibnamefont {Drinkwater}}, \ and\ \bibinfo
		{author} {\bibfnamefont {J.~D.}\ \bibnamefont {Barrow}},\ }\href {\doibase
		10.1103/PhysRevLett.82.884} {\bibfield  {journal} {\bibinfo  {journal} {Phys.
				Rev. Lett.}\ }\textbf {\bibinfo {volume} {82}},\ \bibinfo {pages} {884}
		(\bibinfo {year} {1999})}\BibitemShut {NoStop}%
	\bibitem [{\citenamefont {Hilger}\ \emph {et~al.}(2017)\citenamefont {Hilger},
		\citenamefont {{Gomez-Rocha}},\ and\ \citenamefont
		{Krassnigg}}]{Hilger2015EPJC}%
	\BibitemOpen
	\bibfield  {author} {\bibinfo {author} {\bibfnamefont {T.}~\bibnamefont
			{Hilger}}, \bibinfo {author} {\bibfnamefont {M.}~\bibnamefont
			{{Gomez-Rocha}}}, \ and\ \bibinfo {author} {\bibfnamefont {A.}~\bibnamefont
			{Krassnigg}},\ }\href {\doibase 10.1140/epjc/s10052-017-5163-4} {\bibfield
		{journal} {\bibinfo  {journal} {Eur. Phys. J. C}\ }\textbf {\bibinfo {volume}
			{77}},\ \bibinfo {pages} {625} (\bibinfo {year} {2017})}\BibitemShut
	{NoStop}%
	\bibitem [{\citenamefont {Bhagwat}\ \emph {et~al.}(2007)\citenamefont
		{Bhagwat}, \citenamefont {Krassnigg}, \citenamefont {Maris},\ and\
		\citenamefont {Roberts}}]{Bhagwat2007EPJA}%
	\BibitemOpen
	\bibfield  {author} {\bibinfo {author} {\bibfnamefont {M.~S.}\ \bibnamefont
			{Bhagwat}}, \bibinfo {author} {\bibfnamefont {A.}~\bibnamefont {Krassnigg}},
		\bibinfo {author} {\bibfnamefont {P.}~\bibnamefont {Maris}}, \ and\ \bibinfo
		{author} {\bibfnamefont {C.~D.}\ \bibnamefont {Roberts}},\ }\href {\doibase
		10.1140/epja/i2006-10271-9} {\bibfield  {journal} {\bibinfo  {journal} {Eur.
				Phys. J. A}\ }\textbf {\bibinfo {volume} {31}},\ \bibinfo {pages} {630}
		(\bibinfo {year} {2007})}\BibitemShut {NoStop}%
	\bibitem [{\citenamefont {Xing}\ \emph {et~al.}(2021)\citenamefont {Xing},
		\citenamefont {Chao}, \citenamefont {Chang},\ and\ \citenamefont
		{Liu}}]{Xing2021arXiv}%
	\BibitemOpen
	\bibfield  {author} {\bibinfo {author} {\bibfnamefont {Z.}~\bibnamefont
			{Xing}}, \bibinfo {author} {\bibfnamefont {J.}~\bibnamefont {Chao}}, \bibinfo
		{author} {\bibfnamefont {L.}~\bibnamefont {Chang}}, \ and\ \bibinfo {author}
		{\bibfnamefont {Y.-X.}\ \bibnamefont {Liu}},\ }\href@noop {} {\  (\bibinfo
		{year} {2021})},\ \Eprint {http://arxiv.org/abs/2110.01245}
	{arXiv:2110.01245} \BibitemShut {NoStop}%
	\bibitem [{\citenamefont {Macfarlane}(1962)}]{Macfarlane1962RMP}%
	\BibitemOpen
	\bibfield  {author} {\bibinfo {author} {\bibfnamefont {A.~J.}\ \bibnamefont
			{Macfarlane}},\ }\href {\doibase 10.1103/RevModPhys.34.41} {\bibfield
		{journal} {\bibinfo  {journal} {Rev. Mod. Phys.}\ }\textbf {\bibinfo {volume}
			{34}},\ \bibinfo {pages} {41} (\bibinfo {year} {1962})}\BibitemShut {NoStop}%
	\bibitem [{\citenamefont {Eichmann}\ \emph {et~al.}(2016)\citenamefont
		{Eichmann}, \citenamefont {{Sanchis-Alepuz}}, \citenamefont {Williams},
		\citenamefont {Alkofer},\ and\ \citenamefont {Fischer}}]{Eichmann2016PPNP}%
	\BibitemOpen
	\bibfield  {author} {\bibinfo {author} {\bibfnamefont {G.}~\bibnamefont
			{Eichmann}}, \bibinfo {author} {\bibfnamefont {H.}~\bibnamefont
			{{Sanchis-Alepuz}}}, \bibinfo {author} {\bibfnamefont {R.}~\bibnamefont
			{Williams}}, \bibinfo {author} {\bibfnamefont {R.}~\bibnamefont {Alkofer}}, \
		and\ \bibinfo {author} {\bibfnamefont {C.~S.}\ \bibnamefont {Fischer}},\
	}\href {\doibase 10.1016/j.ppnp.2016.07.001} {\bibfield  {journal} {\bibinfo
			{journal} {Prog. Part. Nucl. Phys.}\ }\textbf {\bibinfo {volume} {91}},\
		\bibinfo {pages} {1} (\bibinfo {year} {2016})}\BibitemShut {NoStop}%
	\bibitem [{\citenamefont {Williams}\ \emph {et~al.}(2007)\citenamefont
		{Williams}, \citenamefont {Fischer},\ and\ \citenamefont
		{Pennington}}]{Williams2006PLB}%
	\BibitemOpen
	\bibfield  {author} {\bibinfo {author} {\bibfnamefont {R.}~\bibnamefont
			{Williams}}, \bibinfo {author} {\bibfnamefont {C.}~\bibnamefont {Fischer}}, \
		and\ \bibinfo {author} {\bibfnamefont {M.}~\bibnamefont {Pennington}},\
	}\href {\doibase 10.1016/j.physletb.2006.12.055} {\bibfield  {journal}
		{\bibinfo  {journal} {Phys. Lett. B}\ }\textbf {\bibinfo {volume} {645}},\
		\bibinfo {pages} {167} (\bibinfo {year} {2007})}\BibitemShut {NoStop}%
	\bibitem [{\citenamefont {Braun}\ \emph {et~al.}(2020)\citenamefont {Braun},
		\citenamefont {Fu}, \citenamefont {Pawlowski}, \citenamefont {Rennecke},
		\citenamefont {Rosenbl{\"u}h},\ and\ \citenamefont {Yin}}]{Braun2020PRD}%
	\BibitemOpen
	\bibfield  {author} {\bibinfo {author} {\bibfnamefont {J.}~\bibnamefont
			{Braun}}, \bibinfo {author} {\bibfnamefont {W.-J.}\ \bibnamefont {Fu}},
		\bibinfo {author} {\bibfnamefont {J.~M.}\ \bibnamefont {Pawlowski}}, \bibinfo
		{author} {\bibfnamefont {F.}~\bibnamefont {Rennecke}}, \bibinfo {author}
		{\bibfnamefont {D.}~\bibnamefont {Rosenbl{\"u}h}}, \ and\ \bibinfo {author}
		{\bibfnamefont {S.}~\bibnamefont {Yin}},\ }\href {\doibase
		10.1103/PhysRevD.102.056010} {\bibfield  {journal} {\bibinfo  {journal}
			{Phys. Rev. D}\ }\textbf {\bibinfo {volume} {102}},\ \bibinfo {pages}
		{056010} (\bibinfo {year} {2020})}\BibitemShut {NoStop}%
	\bibitem [{\citenamefont {Bazavov}\ and\ \citenamefont
		{et~al.}(2012)}]{Bazavov2011PRD}%
	\BibitemOpen
	\bibfield  {author} {\bibinfo {author} {\bibfnamefont {A.}~\bibnamefont
			{Bazavov}}\ and\ \bibinfo {author} {\bibnamefont {et~al.}} (\bibinfo
		{collaboration} {HotQCD}),\ }\href {\doibase 10.1103/PhysRevD.85.054503}
	{\bibfield  {journal} {\bibinfo  {journal} {Phys. Rev. D}\ }\textbf {\bibinfo
			{volume} {85}},\ \bibinfo {pages} {054503} (\bibinfo {year}
		{2012})}\BibitemShut {NoStop}%
	\bibitem [{\citenamefont {Gao}\ \emph {et~al.}(2021)\citenamefont {Gao},
		\citenamefont {Papavassiliou},\ and\ \citenamefont {Pawlowski}}]{Gao2021PRD}%
	\BibitemOpen
	\bibfield  {author} {\bibinfo {author} {\bibfnamefont {F.}~\bibnamefont
			{Gao}}, \bibinfo {author} {\bibfnamefont {J.}~\bibnamefont {Papavassiliou}},
		\ and\ \bibinfo {author} {\bibfnamefont {J.~M.}\ \bibnamefont {Pawlowski}},\
	}\href {\doibase doi.org/10.1103/PhysRevD.103.094013} {\bibfield  {journal}
		{\bibinfo  {journal} {Phys. Rev. D}\ }\textbf {\bibinfo {volume} {103}},\
		\bibinfo {pages} {094013} (\bibinfo {year} {2021})}\BibitemShut {NoStop}%
	\bibitem [{\citenamefont {Ren}\ \emph {et~al.}(2015)\citenamefont {Ren},
		\citenamefont {Geng},\ and\ \citenamefont {Meng}}]{Ren2015PRD}%
	\BibitemOpen
	\bibfield  {author} {\bibinfo {author} {\bibfnamefont {X.-L.}\ \bibnamefont
			{Ren}}, \bibinfo {author} {\bibfnamefont {L.-S.}\ \bibnamefont {Geng}}, \
		and\ \bibinfo {author} {\bibfnamefont {J.}~\bibnamefont {Meng}},\ }\href
	{\doibase 10.1103/PhysRevD.91.051502} {\bibfield  {journal} {\bibinfo
			{journal} {Phys. Rev. D}\ }\textbf {\bibinfo {volume} {91}},\ \bibinfo
		{pages} {051502} (\bibinfo {year} {2015})}\BibitemShut {NoStop}%
	\bibitem [{\citenamefont {Feynman}(1939)}]{Feynman1939PR}%
	\BibitemOpen
	\bibfield  {author} {\bibinfo {author} {\bibfnamefont {R.~P.}\ \bibnamefont
			{Feynman}},\ }\href {\doibase 10.1103/PhysRev.56.340} {\bibfield  {journal}
		{\bibinfo  {journal} {Phys. Rev.}\ }\textbf {\bibinfo {volume} {56}},\
		\bibinfo {pages} {340} (\bibinfo {year} {1939})}\BibitemShut {NoStop}%
	\bibitem [{\citenamefont {Munczek}(1995)}]{Munczek1995PRD}%
	\BibitemOpen
	\bibfield  {author} {\bibinfo {author} {\bibfnamefont {H.~J.}\ \bibnamefont
			{Munczek}},\ }\href {\doibase 10.1103/PhysRevD.52.4736} {\bibfield  {journal}
		{\bibinfo  {journal} {Phys. Rev. D}\ }\textbf {\bibinfo {volume} {52}},\
		\bibinfo {pages} {4736} (\bibinfo {year} {1995})}\BibitemShut {NoStop}%
	\bibitem [{\citenamefont {Bender}\ \emph {et~al.}(1996)\citenamefont {Bender},
		\citenamefont {Roberts},\ and\ \citenamefont {Smekal}}]{Bender1996PLB}%
	\BibitemOpen
	\bibfield  {author} {\bibinfo {author} {\bibfnamefont {A.}~\bibnamefont
			{Bender}}, \bibinfo {author} {\bibfnamefont {C.}~\bibnamefont {Roberts}}, \
		and\ \bibinfo {author} {\bibfnamefont {L.}~\bibnamefont {Smekal}},\ }\href
	{\doibase https://doi.org/10.1016/0370-2693(96)00372-3} {\bibfield  {journal}
		{\bibinfo  {journal} {Phys. Lett. B}\ }\textbf {\bibinfo {volume} {380}},\
		\bibinfo {pages} {7} (\bibinfo {year} {1996})}\BibitemShut {NoStop}%
	\bibitem [{\citenamefont {Qin}\ \emph {et~al.}(2011{\natexlab{b}})\citenamefont
		{Qin}, \citenamefont {Chang}, \citenamefont {Liu}, \citenamefont {Roberts},\
		and\ \citenamefont {Wilson}}]{Qin2011PRC}%
	\BibitemOpen
	\bibfield  {author} {\bibinfo {author} {\bibfnamefont {S.-X.}\ \bibnamefont
			{Qin}}, \bibinfo {author} {\bibfnamefont {L.}~\bibnamefont {Chang}}, \bibinfo
		{author} {\bibfnamefont {Y.-X.}\ \bibnamefont {Liu}}, \bibinfo {author}
		{\bibfnamefont {C.~D.}\ \bibnamefont {Roberts}}, \ and\ \bibinfo {author}
		{\bibfnamefont {D.~J.}\ \bibnamefont {Wilson}},\ }\href {\doibase
		10.1103/PhysRevC.84.042202} {\bibfield  {journal} {\bibinfo  {journal} {Phys.
				Rev. C}\ }\textbf {\bibinfo {volume} {84}},\ \bibinfo {pages} {042202}
		(\bibinfo {year} {2011}{\natexlab{b}})}\BibitemShut {NoStop}%
	\bibitem [{\citenamefont {Maris}\ and\ \citenamefont
		{Tandy}(1999)}]{Maris1999PRC}%
	\BibitemOpen
	\bibfield  {author} {\bibinfo {author} {\bibfnamefont {P.}~\bibnamefont
			{Maris}}\ and\ \bibinfo {author} {\bibfnamefont {P.~C.}\ \bibnamefont
			{Tandy}},\ }\href {\doibase 10.1103/PhysRevC.60.055214} {\bibfield  {journal}
		{\bibinfo  {journal} {Phys. Rev. C}\ }\textbf {\bibinfo {volume} {60}},\
		\bibinfo {pages} {055214} (\bibinfo {year} {1999})}\BibitemShut {NoStop}%
	\bibitem [{\citenamefont {Chang}\ \emph {et~al.}(2021)\citenamefont {Chang},
		\citenamefont {Liu}, \citenamefont {Raya}, \citenamefont
		{{Rodr{\'i}guez-Quintero}},\ and\ \citenamefont {Yang}}]{Chang2021PRD}%
	\BibitemOpen
	\bibfield  {author} {\bibinfo {author} {\bibfnamefont {L.}~\bibnamefont
			{Chang}}, \bibinfo {author} {\bibfnamefont {Y.-B.}\ \bibnamefont {Liu}},
		\bibinfo {author} {\bibfnamefont {K.}~\bibnamefont {Raya}}, \bibinfo {author}
		{\bibfnamefont {J.}~\bibnamefont {{Rodr{\'i}guez-Quintero}}}, \ and\ \bibinfo
		{author} {\bibfnamefont {Y.-B.}\ \bibnamefont {Yang}},\ }\href {\doibase
		10.1103/PhysRevD.104.094509} {\bibfield  {journal} {\bibinfo  {journal}
			{Phys. Rev. D}\ }\textbf {\bibinfo {volume} {104}},\ \bibinfo {pages}
		{094509} (\bibinfo {year} {2021})}\BibitemShut {NoStop}%
	\bibitem [{\citenamefont {Qin}\ and\ \citenamefont
		{Roberts}(2021)}]{Qin2021CPL}%
	\BibitemOpen
	\bibfield  {author} {\bibinfo {author} {\bibfnamefont {S.-X.}\ \bibnamefont
			{Qin}}\ and\ \bibinfo {author} {\bibfnamefont {C.~D.}\ \bibnamefont
			{Roberts}},\ }\href {\doibase 10.1088/0256-307X/38/7/071201} {\bibfield
		{journal} {\bibinfo  {journal} {Chin. Phys. Lett.}\ }\textbf {\bibinfo
			{volume} {38}},\ \bibinfo {pages} {071201} (\bibinfo {year}
		{2021})}\BibitemShut {NoStop}%
	\bibitem [{\citenamefont {Qin}\ \emph {et~al.}(2020)\citenamefont {Qin},
		\citenamefont {Qin},\ and\ \citenamefont {Liu}}]{Qp2019PRD}%
	\BibitemOpen
	\bibfield  {author} {\bibinfo {author} {\bibfnamefont {P.}~\bibnamefont
			{Qin}}, \bibinfo {author} {\bibfnamefont {S.-X.}\ \bibnamefont {Qin}}, \ and\
		\bibinfo {author} {\bibfnamefont {Y.-X.}\ \bibnamefont {Liu}},\ }\href
	{\doibase 10.1103/PhysRevD.101.114014} {\bibfield  {journal} {\bibinfo
			{journal} {Phys. Rev. D}\ }\textbf {\bibinfo {volume} {101}},\ \bibinfo
		{pages} {114014} (\bibinfo {year} {2020})}\BibitemShut {NoStop}%
	\bibitem [{\citenamefont {Cloët}\ and\ \citenamefont
		{Roberts}(2014)}]{Cloet2014PPNP}%
	\BibitemOpen
	\bibfield  {author} {\bibinfo {author} {\bibfnamefont {I.~C.}\ \bibnamefont
			{Cloët}}\ and\ \bibinfo {author} {\bibfnamefont {C.~D.}\ \bibnamefont
			{Roberts}},\ }\href {\doibase https://doi.org/10.1016/j.ppnp.2014.02.001}
	{\bibfield  {journal} {\bibinfo  {journal} {Prog. Part. Nucl. Phys.}\
		}\textbf {\bibinfo {volume} {77}},\ \bibinfo {pages} {1} (\bibinfo {year}
		{2014})}\BibitemShut {NoStop}%
	\bibitem [{\citenamefont {Chang}\ \emph {et~al.}(2011)\citenamefont {Chang},
		\citenamefont {Roberts},\ and\ \citenamefont {Tandy}}]{Chang2011CJP}%
	\BibitemOpen
	\bibfield  {author} {\bibinfo {author} {\bibfnamefont {L.}~\bibnamefont
			{Chang}}, \bibinfo {author} {\bibfnamefont {C.~D.}\ \bibnamefont {Roberts}},
		\ and\ \bibinfo {author} {\bibfnamefont {P.~C.}\ \bibnamefont {Tandy}},\
	}\href@noop {} {\bibfield  {journal} {\bibinfo  {journal} {Chin. J. Phys.}\
		}\textbf {\bibinfo {volume} {49}},\ \bibinfo {pages} {955} (\bibinfo {year}
		{2011})}\BibitemShut {NoStop}%
	\bibitem [{\citenamefont {Qin}\ \emph {et~al.}(2019)\citenamefont {Qin},
		\citenamefont {Roberts},\ and\ \citenamefont {Schmidt}}]{Qin2019FBS}%
	\BibitemOpen
	\bibfield  {author} {\bibinfo {author} {\bibfnamefont {S.-X.}\ \bibnamefont
			{Qin}}, \bibinfo {author} {\bibfnamefont {C.~D.}\ \bibnamefont {Roberts}}, \
		and\ \bibinfo {author} {\bibfnamefont {S.~M.}\ \bibnamefont {Schmidt}},\
	}\href {\doibase 10.1007/s00601-019-1488-x} {\bibfield  {journal} {\bibinfo
			{journal} {Few Body Syst.}\ }\textbf {\bibinfo {volume} {60}},\ \bibinfo
		{pages} {26} (\bibinfo {year} {2019})}\BibitemShut {NoStop}%
	\bibitem [{\citenamefont {Aguilar}\ \emph {et~al.}(2009)\citenamefont
		{Aguilar}, \citenamefont {Binosi}, \citenamefont {Papavassiliou},\ and\
		\citenamefont {Rodr\'{\i}guez-Quintero}}]{Aguilar2009PRD}%
	\BibitemOpen
	\bibfield  {author} {\bibinfo {author} {\bibfnamefont {A.~C.}\ \bibnamefont
			{Aguilar}}, \bibinfo {author} {\bibfnamefont {D.}~\bibnamefont {Binosi}},
		\bibinfo {author} {\bibfnamefont {J.}~\bibnamefont {Papavassiliou}}, \ and\
		\bibinfo {author} {\bibfnamefont {J.}~\bibnamefont
			{Rodr\'{\i}guez-Quintero}},\ }\href {\doibase 10.1103/PhysRevD.80.085018}
	{\bibfield  {journal} {\bibinfo  {journal} {Phys. Rev. D}\ }\textbf {\bibinfo
			{volume} {80}},\ \bibinfo {pages} {085018} (\bibinfo {year}
		{2009})}\BibitemShut {NoStop}%
	\bibitem [{\citenamefont {Matevosyan}\ \emph {et~al.}(2007)\citenamefont
		{Matevosyan}, \citenamefont {Thomas},\ and\ \citenamefont
		{Tandy}}]{Matevosyan2007PRC}%
	\BibitemOpen
	\bibfield  {author} {\bibinfo {author} {\bibfnamefont {H.~H.}\ \bibnamefont
			{Matevosyan}}, \bibinfo {author} {\bibfnamefont {A.~W.}\ \bibnamefont
			{Thomas}}, \ and\ \bibinfo {author} {\bibfnamefont {P.~C.}\ \bibnamefont
			{Tandy}},\ }\href {\doibase 10.1103/PhysRevC.75.045201} {\bibfield  {journal}
		{\bibinfo  {journal} {Phys. Rev. C}\ }\textbf {\bibinfo {volume} {75}},\
		\bibinfo {pages} {045201} (\bibinfo {year} {2007})}\BibitemShut {NoStop}%
	\bibitem [{\citenamefont {Chang}\ and\ \citenamefont
		{Roberts}(2009)}]{Chang2009PRL}%
	\BibitemOpen
	\bibfield  {author} {\bibinfo {author} {\bibfnamefont {L.}~\bibnamefont
			{Chang}}\ and\ \bibinfo {author} {\bibfnamefont {C.~D.}\ \bibnamefont
			{Roberts}},\ }\href {\doibase 10.1103/PhysRevLett.103.081601} {\bibfield
		{journal} {\bibinfo  {journal} {Phys. Rev. Lett.}\ }\textbf {\bibinfo
			{volume} {103}},\ \bibinfo {pages} {081601} (\bibinfo {year}
		{2009})}\BibitemShut {NoStop}%
	\bibitem [{\citenamefont {Chang}\ and\ \citenamefont
		{Roberts}(2012)}]{Chang2012PRC85}%
	\BibitemOpen
	\bibfield  {author} {\bibinfo {author} {\bibfnamefont {L.}~\bibnamefont
			{Chang}}\ and\ \bibinfo {author} {\bibfnamefont {C.~D.}\ \bibnamefont
			{Roberts}},\ }\href {\doibase 10.1103/PhysRevC.85.052201} {\bibfield
		{journal} {\bibinfo  {journal} {Phys. Rev. C}\ }\textbf {\bibinfo {volume}
			{85}},\ \bibinfo {pages} {052201} (\bibinfo {year} {2012})}\BibitemShut
	{NoStop}%
	\bibitem [{\citenamefont {Binosi}\ \emph {et~al.}(2016)\citenamefont {Binosi},
		\citenamefont {Chang}, \citenamefont {Papavassiliou}, \citenamefont {Qin},\
		and\ \citenamefont {Roberts}}]{Binosi2016PRD}%
	\BibitemOpen
	\bibfield  {author} {\bibinfo {author} {\bibfnamefont {D.}~\bibnamefont
			{Binosi}}, \bibinfo {author} {\bibfnamefont {L.}~\bibnamefont {Chang}},
		\bibinfo {author} {\bibfnamefont {J.}~\bibnamefont {Papavassiliou}}, \bibinfo
		{author} {\bibfnamefont {S.-X.}\ \bibnamefont {Qin}}, \ and\ \bibinfo
		{author} {\bibfnamefont {C.~D.}\ \bibnamefont {Roberts}},\ }\href {\doibase
		10.1103/PhysRevD.93.096010} {\bibfield  {journal} {\bibinfo  {journal} {Phys.
				Rev. D}\ }\textbf {\bibinfo {volume} {93}},\ \bibinfo {pages} {096010}
		(\bibinfo {year} {2016})}\BibitemShut {NoStop}%
\end{thebibliography}

\end{document}